\documentclass[12pt]{article}
\usepackage{amssymb,amsmath,epsfig}
\usepackage[cp1251]{inputenc}
\allowdisplaybreaks

\begin{document}

\title{\bf Compact Stars in $f(\mathcal{R,G,T})$ Gravity}
\author{M. Ilyas \thanks{ilyas\_mia@yahoo.com}\\
Institute of Physics, Gomal University,\\
Dera Ismail Khan, 29220, Khyber Pakhtunkhwa, Pakistan}

\date{}
\maketitle

\begin{abstract}
The present work is to introduce a new kind of modified gravitational theory, named as $f(\mathcal{R,G,T})$ (also $f(\mathcal{R,T,G})$) gravity, where $\mathcal{R}$ is the Ricci scalar, $\mathcal{G}$ is Gauss-Bonnet invariant and $\mathcal{T}$ is the trace of the energy-momentum tensor. With the help of different models in this gravity, we investigate some physical features of different relativistic compact stars. For this purpose, we develop the effectively modified field equations, conservation equation, and the equation of motion for test particle. Then, we check the impact of additional force (massive test particle followed by a non-geodesic line of geometry) on compact objects. Furthermore, we took three notable stars named as $Her X-1$, $SAXJ1808.4-3658$ and $4U1820-30$. The physical behavior of the energy density, anisotropic pressures, different energy conditions, stability, anisotropy, and the equilibrium scenario of these strange compact stars are analyzed through various plots. Finally, we conclude that the energy conditions hold, and the core of these stars is so dense.
\end{abstract}

\section{Introduction}

As the expansion of the universe is accelerating and has been confirmed by various sets of experimental data getting from the astronomical scenario, like supernova type Ia, cosmic microwave background radiation, massive structures, and many more \cite{1}. The well-known fact is that general relativity (GR) is insufficient to explain the current expansion of the universe and this accelerated expansion is due to the result of an unknown force dubbed as dark energy (DE), which is negatively pressurized. For the investigation of DE, different Modified theories of gravity (MTGs) are supposed to be a strong candidate which unveil the hidden secrets of the universe.  These MTGs are normally established by modifying the Einstein-Hilbert action.
Some MTGs are, such that $f(\cal{R})$ gravity theory \cite{615} in which ricci scalar $\cal{R}$ has been replaced by an arbitrary function of $\cal{R}$ (like $f(\cal{R})$) in Einstein-Hilbert action. The accelerated expansion of the universe is well explained by $f(\cal{R})$ gravity \cite{616} and the first more consistent theory which satisfies newton law was proposed in \cite{new00} and also derived the conditions for a viable cosmological models \cite{617}.\\
A simple and general viable model in modified $f(\cal{R})$ gravity were proposed \cite{new0} in which a non-minimal coupling term was used between the matter and geometry part. This coupling term is treated, as the source of gravitation, which explains the current accelerated expansion of the universe. For further review, one can see Refs. \cite{618}.
An extra force arises due to this coupling, considering the different forms of matter in action (Lagrangian density), yields the extra force \cite{619} and it can be easily seen that the more suitable and natural form of action doesn't imply to vanish this additional force \cite{6110}. The effects of the non-minimal coupling over the relativistic stellar equilibrium case were also studied.\\
The most interesting MTG is the Gauss-Bonnet $(GB)$ gravity.  The GB invariant term is written as
\begin{equation}\nonumber
\mathcal{G}=\mathcal{R}_{\alpha\beta\xi\eta}
\mathcal{R}^{\alpha\beta\xi\eta}-4\mathcal{R}_{\beta\xi}R^{\beta\xi}+\mathcal{R}^2,
\end{equation}
here $(\mathcal{G})$ is called a GB invariant and $ \mathcal{R}$ is the Ricci scalar, $ \mathcal{R}_{\beta\xi}$ is Ricci tensor and $\mathcal{R}_{\alpha\beta\xi\eta}$ represent the Riemann tensor. GB theory is free from spin two ghosts instabilities \cite{3} and is the 2nd order Lovelock scalar invariant.\\
As GB term is a 4D topological invariant and have no impact on the field equations but offers some exciting results if its coupled with a scalar field or written in the form of any function $f(\mathcal{G})$ \cite{4}.
The arbitrary function's approach is suggested by Nojiri and Odintsov and is known as modified $f(\mathcal{G})$ gravity theory \cite{4a}. In a similar fashion, as like other MTGs, to study DE, this theory is a good alternative which also incorporates the solar system constraints \cite{5}.
In this framework, the author discussed the phase transition, from deceleration to accelerated and non-phantom to phantom, and early and late-time unification of expansion of the universe \cite{6}.
With the help of MTGs, the captivating problem regarding the accelerated expansion of the universe is well addressed.\\

The mathematical modeling of compact objects like neutrons stars, pulsars, and black holes attract researchers, recently, with the help of different MTGs, many aspects of different strange stars were investigated \cite{ref19,ref19a,ref19b,ref19c}. To get realistic mathematical modeling, the spherical symmetric geometry is very fruitful having choices in matter distribution. In the study of compacts objects, many researchers took the distribution of matter as a perfect fluid while the anisotropic fluid is also considered. As compared to a perfect fluid, the stability is a little bit disturbed due to the anisotropy. With the help of the equation of state, the effects of local anisotropy are studied \cite{ref21}. Hence, it seems more appropriate to assume the anisotropic matter content with MTG models. Many of the physical features of different compacts stars were already discussed in the presence of anisotropic fluid and charge \cite{ref25,ref26,ref27,ref28}.\\

Here in this research work, we present a new kind of MTG labeled, $f(\mathcal{R,G,T})$ theory of gravity, in which we obtained the gravitational Lagrangian by the addition of a generic function $f(\mathcal{R,G,T})$ instead of $\cal{R}$ in the Hilbert Einstein action.  This paper having the format as follows, we formulate the field equation, motion of test particle, and assume the anisotropic form of matter distribution with static spherical geometry in section 2. The next section is devoted to analyzing some stable geometries, in order to make our theory more applicable, we assume different compact stars as stable geometries and studied some features of these stars like stability, variation in energy density, stresses, energy conditions, and many more. We summarize our results and discussion in the final section.

\section{$f(\mathcal{R,G,T})$ theory of gravity}

In order to established $f(\mathcal{R,G,T})$ theory of gravity and to devolve its field equations, we let the following form of Action
\begin{equation}\label{1}
\mathcal{S}_{f(\mathcal{R,G,T})}=\frac{\kappa^{-2}}{2}\int d^{4}x\sqrt{-g}[f(\mathcal{R,G,T})+\mathcal{L}_{m}],
\end{equation}
Here $\kappa$ represent coupling constant and $g$ represent the determinant of the metric tensor $(g_{\alpha\beta})$. The
energy momentum tensor, $\mathcal{T}_{\gamma\delta}$, is written as \cite{16}
\begin{equation}\label{2}
\mathcal{T}_{\gamma\delta}=-\frac{2}{\sqrt{-g}}\frac{\delta(\mathcal{L}_{m}\sqrt{-g})}{\delta
g^{\gamma\delta}}.
\end{equation}
which further equals to
\begin{equation}\label{3}
\mathcal{T}_{\gamma\delta}=g_{\gamma\delta
}\mathcal{L}_{m}-2\frac{\partial\mathcal{L}_{m}}{\partial
g^{\gamma\delta }}.
\end{equation}
The variation in the action written in Eq. (\ref{1}) gives
\begin{eqnarray}\label{4}\nonumber
\delta {\cal S} &= \frac{{{\kappa ^{ - 2}}}}{2}\int {{d^4}} x[(f({\cal R},{\cal G},{\cal T})\delta \sqrt { - g}  + \sqrt { - g} ({f_{\cal R}}({\cal R},{\cal G},{\cal T})\delta {\cal R}\\
& + {f_{\cal G}}({\cal R},{\cal G},{\cal T})\delta {\cal G} + {f_{\cal T}}({\cal R},{\cal G},{\cal T})\delta {\cal T}) + \sqrt { - g} \delta {{\cal L}_m}] = 0,
\end{eqnarray}
where $f_{\mathcal{R}}(\mathcal{R,G,T})=\frac{\partial
f(\mathcal{R,G,T})}{\partial\mathcal{R}}$,
$f_{\mathcal{G}}(\mathcal{R,G,T})=\frac{\partial
f(\mathcal{R,G,T})}{\partial\mathcal{G}}$ and
$f_{\mathcal{T}}(\mathcal{R,G,T})=\frac{\partial f(\mathcal{R,G,T})}{\partial \mathcal{T}}$.
The variation of $\sqrt{-g},~\mathcal{R}^{\xi}_{\alpha\beta\eta},~\mathcal{R}_{\alpha\eta}$ and $\mathcal{R}$
give the following expressions
\begin{eqnarray}\nonumber
\delta\sqrt{-g}&=&-\frac{1}{2}\sqrt{-g}g_{\rho\sigma}\delta
g^{\rho\sigma },\\\nonumber\delta
\mathcal{R}^{\xi}_{\rho\sigma\eta}&=&\nabla_{\sigma}(\delta\Gamma^{\xi}_{\eta\rho})
-\nabla_{\eta}(\delta\Gamma^{\xi}_{\sigma\rho}),\\\nonumber&=&(g_{\rho\lambda}
\nabla_{[\eta}\nabla_{\sigma
]}+g_{\lambda[\sigma}\nabla_{\eta]}\nabla_{\rho}) \delta
g^{\xi\lambda}+\nabla_{[\eta}\nabla^{\xi}\delta
g_{\sigma]\rho},\\\label{4a}\delta R_{\rho\eta}&=&\delta
\mathcal{R}^{\xi}_{\rho\xi\eta},\quad\delta
\mathcal{R}=(\mathcal{R}_{\rho\sigma}+g_{\rho\sigma}\nabla^2-\nabla_{\rho}\nabla_{\sigma})
\delta g^{\rho\sigma},
\end{eqnarray}
where $\nabla_{\rho}$ and $\Gamma^{\xi}_{\rho\sigma}$ represent
the covariant-derivative and Christoffel symbols, respectively. The
variation of $\mathcal{R}$, $\mathcal{G}$ and $\mathcal{T}$ yield
\begin{eqnarray}\nonumber
\delta \mathcal{R} &=& {\mathcal{R}_{\rho \sigma }}\delta {g^{\rho \sigma }} - {\nabla _\rho }{\nabla _\sigma }\delta {g^{\rho \sigma }} - {g_{\rho \sigma }}\Box\delta {g^{\rho \sigma }}\\\nonumber
\delta\mathcal{G}&=&2\mathcal{R}\delta
\mathcal{R}-4\delta(R_{\rho\sigma}\mathcal{R}^{\rho\sigma})+\delta(\mathcal{R}_{\rho
\sigma\xi\eta}R^{\rho\sigma\xi\eta}),
\\\label{4b}\delta \mathcal{T}&=&(\mathcal{T}_{\rho\sigma}+\Theta_{\rho\sigma})\delta
g^{\rho\sigma},\quad\Theta_{\rho\sigma}=g^{\xi\eta}\frac{\delta
\mathcal{T}_{\xi\eta}}{\delta g_{\rho\sigma}}.
\end{eqnarray}

Using Eq.(\ref{4}) along with the help of these above relations, we get the desire
field equations of the $f(\mathcal{R,G,T})$ gravity, as follows

\begin{align}\nonumber
&{\mathcal{G}_{\rho \sigma }} = \\\nonumber\label{5}
&\frac{1}{{{f_{\cal R}}\left( {{\cal R},{\cal G},{\cal T}} \right)}}\left[ {{\kappa ^2}{{\cal T}_{\rho \sigma }}} \right. - \left( {{{\cal T}_{\rho \sigma }} + {\Theta _{\rho \sigma }}} \right){f_T}\left( {{\cal R},{\cal G},{\cal T}} \right) \\\nonumber
& + \frac{1}{2}{g_{\rho \sigma }}(f\left( {{\cal R},{\cal G},{\cal T}} \right) + \mathcal{R}{f_{\cal R}}\left( {{\cal R},{\cal G},{\cal T}} \right)) + {\nabla _\rho }{\nabla _\sigma }{f_{\cal R}}\left( {{\cal R},{\cal G},{\cal T}} \right)\\\nonumber
& - {g_{\rho \sigma }}\Box{f_{\cal R}}\left( {{\cal R},{\cal G},{\cal T}} \right) - (2\mathcal{R}{\mathcal{R}_{\rho \sigma }} - 4\mathcal{R}_\rho ^\xi {\mathcal{R}_{\xi \sigma }} - 4{\mathcal{R}_{\rho \xi \sigma \eta }}{\mathcal{R}^{\xi \eta }}\\\nonumber
& + 2\mathcal{R}_\rho ^{\xi \eta \delta }{R_{\sigma \xi \eta \delta }}){f_{\cal G}}\left( {{\cal R},{\cal G},{\cal T}} \right) - (2\mathcal{R}{g_{\rho \sigma }}{\nabla ^2} - 2\mathcal{R}{\nabla _\rho }{\nabla _\sigma } - 4{g_{\rho \sigma }}{\mathcal{R}^{\xi \eta }}{\nabla _\xi }{\nabla _\eta }\\
& - 4{\mathcal{R}_{\rho \sigma }}{\nabla ^2} + 4\mathcal{R}_\rho ^\xi {\nabla _\sigma }{\nabla _\xi } + 4\mathcal{R}_\sigma ^\xi {\nabla _\rho }{\nabla _\xi } + 4{\mathcal{R}_{\rho \xi \sigma \eta }}{\nabla ^\xi }{\nabla ^\eta })\left. {{f_{\cal G}}\left( {{\cal R},{\cal G},{\cal T}} \right)} \right],
\end{align}
where $\mathcal{G}_{\rho\sigma}=\mathcal{R}_{\rho\sigma}-\frac{1}{2}g_{\rho\sigma}\mathcal{R}$ is the
Einstein tensor. \\
It is worth mentioning here that for $f(\mathcal{R,G,T})$ can be easily reduces to the following modified theories e.g.
$$f({\cal R},{\cal G},{\cal T}) \equiv \left. {\left| {\begin{array}{*{20}{c}}
G.R\\
{f({\cal R})}\\
{f({\cal R},{\cal T})}\\
{f({\cal G})}\\
{f({\cal R},{\cal G})}\\
{f({\cal G},{\cal T})}
\end{array}} \right.} \right\}$$
We can get the field equations of these theories from Eq. (\ref{5}). Furthermore, the general relativity field
equations can easily archived back by $f(\mathcal{R,G,T})=\mathcal{R}$ and trace of
equation (\ref{5}) can be written as
\begin{align}\nonumber
&{\kappa ^2}{\cal T} - ({\cal T} + \Theta ){f_{\cal T}}({\cal R},{\cal G},{\cal T})  - {f_{\cal R}}({\cal R},{\cal G},{\cal T})R - 3\Box {f_{\cal R}}({\cal R},{\cal G},{\cal T})+ 2{\cal G}{f_{\cal G}}({\cal R},{\cal G},{\cal T})\\
&- 2\mathcal{R}{\nabla ^2}{f_{\cal G}}({\cal R},{\cal G},{\cal T}) + 4{R^{\rho \sigma }}{\nabla _\rho }{\nabla _\sigma }{f_{\cal G}}({\cal R},{\cal G},{\cal T})+ 2f({\cal R},{\cal G},{\cal T}) = 0,
\end{align}
where $\Theta=\Theta^{\rho}_{\rho}$. Moreover, the divergence of Eq. (\ref{5}) is non-zero, like
\begin{align}\label{5a}\nonumber
{\nabla ^\rho }{{\cal T}_{\rho \sigma }} &= \frac{{{f_T}({\cal R},{\cal G},{\cal T})}}{{{\kappa ^2} - {f_{\cal T}}({\cal R},{\cal G},{\cal T})}}\left[ {\left\{ {{{\cal T}_{\rho \sigma }} + {\Theta _{\rho \sigma }}} \right\}{\nabla ^\rho }\left\{ {\ln {f_{\cal T}}({\cal R},{\cal G},{\cal T})} \right\}} \right.\\
& - \left. {\frac{1}{2}{g_{\rho \sigma }}{\nabla ^\rho }{\cal T} + {\nabla ^\rho }{\Theta _{\rho \sigma }}} \right].
\end{align}
Here we have used the following relations
\begin{eqnarray}
&& \left( \nabla_\sigma \Box - \Box \nabla_\sigma \right) \psi = g^{\alpha\beta}
\left( \nabla_\sigma \nabla_\alpha \nabla_\beta - \nabla_\alpha \nabla_\beta
\nabla_\sigma \right) \psi  \notag \\
&=& g^{\alpha\beta} \left( \nabla_\sigma \nabla_\alpha - \nabla_\alpha
\nabla_\sigma \right) \nabla_\beta \psi = g^{\alpha \beta} \mathcal{R}_{\thickspace \beta
\alpha \sigma}^\rho \nabla_\rho \psi  \notag \\
&=& - \mathcal{R}_{\rho \sigma} \nabla^\rho \psi,
\end{eqnarray}
and
\begin{equation}
\nabla_\nu f \left( \mathcal{R,G,T} \right) = f_{\mathcal{R}} \nabla_\nu \mathcal{R} +f_{\mathcal{G}} \nabla_\nu \mathcal{G} + f_{\mathcal{T}} \nabla_\nu \mathcal{T},
\label{eq:nablaf}
\end{equation}
respectively.

In order to get a fruitful expression for $\Theta_{\sigma\rho}$, we need to differentiate Eq.(\ref{3}), as follow
\begin{equation}\label{6}
\frac{\delta \mathcal{T}_{\sigma\rho}}{\delta g^{\xi\eta}}=\frac{\delta
g_{\sigma\rho}}{\delta g^{\xi\eta}}\mathcal{L}_{m}+g_{\sigma\rho}
\frac{\partial\mathcal{L}_{m}}{\partial
g^{\xi\eta}}-2\frac{\partial^2\mathcal{L}_{m}}{\partial
g^{\xi\eta}\partial g^{\sigma\rho}}.
\end{equation}
Using the relations
\begin{equation}\nonumber
\frac{\delta g_{\sigma\rho}}{\delta
g^{\xi\eta}}=-g_{\sigma\mu}g_{\rho\nu}\delta_{\xi\eta}^{\mu\nu},\quad
\delta_{\xi\eta}^{\mu\nu}=\frac{\delta g^{\mu\nu}}{\delta
g^{\xi\eta}},
\end{equation}
where $\delta_{\xi\eta}^{\mu\nu}$ is the generalized Kronecker
symbol. By putting Eq.(\ref{6}) in Eq. (\ref{4b}), we obtain
\begin{equation}\label{7}
\Theta_{\sigma\rho}=-2\mathcal{T}_{\sigma\rho}+g_{\sigma\rho}\mathcal{L}_{m}-2g^{\xi\eta}
\frac{\partial^{2}\mathcal{L}_{m}}{\partial g^{\sigma\rho}\partial
g^{\xi\eta}}.
\end{equation}
This is helpful in finding the tensor $\Theta_{\sigma\rho}$.
\subsection{The equations of motion of test particles in $f(\mathcal{R,G,T})$}

In this subsection, we investigate the motion of a test particle in the presence of
$f(\mathcal{R,G,T})$ gravity. For this, we let the perfect fluid, like
\begin{equation}\label{8}
T_{\sigma\lambda}=(\rho+P)V_{\sigma}V_{\lambda}-P g_{\sigma\lambda},
\end{equation}
Here, $\rho$ is the energy density, $P$ is pressure while $V_{\alpha}$ is four
velocity of fluid. The four velocity obeys the
relation $V_{\sigma}V^{\sigma}=1$ and assume the corresponding Lagrangian
density, $\mathcal{L}_{m}=-P$ \cite{17}. Thus
Eq.(\ref{7}) gives
\begin{equation}\label{9}
\Theta_{\sigma\lambda}=-2\mathcal{T}_{\sigma\lambda}-Pg_{\sigma\lambda}.
\end{equation}
For the divergence of perfect fluid, $\cal{T_{\alpha,\beta}}$, we put Eqs. (\ref{8}) and (\ref{9}) in Eq. (\ref{5a}), we get
\begin{eqnarray}\nonumber
&&\nabla_{\lambda}(\rho+P)V^{\sigma}V^{\lambda}+(\rho+P)[V^{\lambda}\nabla_{\lambda}V^{\sigma}
+V^{\sigma}\nabla_{\lambda}V^{\lambda}]-g^{\sigma\lambda}\nabla_{\lambda}P
\\\nonumber&&=-\frac{2}{3f_{\mathcal{T}}(\mathcal{R,G,T})+2\kappa^2}\left[T^{\sigma\lambda}
\nabla_{\lambda}f_{\mathcal{T}}(\mathcal{R,G,T})+g^{\sigma\lambda}\nabla_{\lambda}(P
f_{\mathcal{T}}(\mathcal{R,G,T}))\right].
\end{eqnarray}
Furthermore, with the help of projection operator
$(h_{\sigma\xi}=g_{\sigma\xi}-V_{\sigma\xi})$, the contraction of above equation give the following
expression
\begin{equation}\label{a}
g_{\sigma\xi}V^{\lambda}\nabla_{\lambda}V^{\sigma}=\frac{(f_{\mathcal{T}}(\mathcal{R,G,T})+2\kappa^2)\nabla_{\lambda}}{(3f_{\mathcal{T}}
(\mathcal{R,G,T}+2\kappa^2)(\rho+P))}h_{\xi}^{\lambda},
\end{equation}
here we have used the relations
$V^{\sigma}\nabla_{\lambda}V_{\sigma}=0,~h_{\sigma\xi}V^{\sigma}=0$
and $h_{\sigma\xi}\mathcal{T}^{\sigma\lambda}=-Ph_{\xi}^{\lambda}$. Now multiplying
the equation (\ref{a}) with $g^{\mu\xi}$ and using the following identity
\cite{17}
\begin{equation}\nonumber
V^{\lambda}\nabla_{\lambda}V^{\sigma}=\frac{d^2x^{\sigma}}{ds^2}+\Gamma^{\sigma}
_{\lambda\xi}V^{\lambda}V^{\xi},
\end{equation}
we reach at the equation of motion for the massive test particle as
\begin{equation}\label{b}
\frac{d^2x^{\sigma}}{ds^2}+\Gamma^{\sigma}
_{\lambda\xi}V^{\lambda}V^{\xi}={\cal{Z}}^{\sigma},
\end{equation}
Here
\begin{equation}\label{c}
{\cal{Z}}^{\sigma}=\frac{(2\kappa^2+f_{\mathcal{T}}(\mathcal{R,G,T}))}{(P+\rho)(3f_{\mathcal{T}}
(\mathcal{R,G,T})+2\kappa^2)}(g^{\sigma\lambda}-V^{\sigma}V^{\lambda})\nabla_{\lambda}P,
\end{equation}
represents the extra additional force (coming from curvature) influence the test particle, which is
perpendicular to the four-velocity $V_{\sigma}$ of the fluid e.g. ${\cal{Z}}^{\sigma}V_{\sigma}=0$. In case of pressureless fluid, $P\approx0$, the Eq. (\ref{c})
give ${\cal{Z}}^{\sigma}=0$ such that the dust particle follow the
geodesic trajectories both in Einstein relativity as well in
$f(\mathcal{R,G,T})$ gravity. In the absence of matter-geometry coupling, the equation of motion for perfect fluid
in Einstein relativity can easily reachable \cite{17a}.

\subsection{The Newtonian limit in $f(\mathcal{R,G,T})$ gravity}\label{The Newtonian limit}

The force term as written in Eq. (\ref{c})  can be formally rewritten as the gradient of a potential $N$ as follows,
\begin{equation}\label{Q}
\frac{(2\kappa^2+f_{\mathcal{T}}(\mathcal{R,G,T}))}{(P+\rho)(3f_{\mathcal{T}}
(\mathcal{R,G,T})+2\kappa^2)}(g^{\sigma\lambda}-V^{\sigma}V^{\lambda})\nabla_{\lambda}P=\nabla _{\sigma }\Big(\log \sqrt{N}\Big)\, ,
\end{equation}
In order to found the geodesic  Eq.~(\ref{b}) from point-particle, let
\begin{equation}
\delta S_{p}=\delta \int L_{p}ds=\delta \int \sqrt{N}\sqrt{g_{\mu \nu
}V^{\mu }V^{\nu }}ds=0\, ,  \label{actpart}
\end{equation}
here $S_{p}$ is the point-like action and $L_{p}=\sqrt{N}\sqrt{g_{\mu \nu }V^{\mu }V^{\nu
}}$ is the point-like Lagrangian for the test particles. Furthermore, the Lagrange equations corresponding to the action~(\ref{actpart}),
\begin{equation}
\frac{d}{ds}\left( \frac{\partial L_{p}}{\partial V^{\lambda }}\right)
 - \frac{\partial L_{p}}{\partial x^{\lambda }}=0\, .
\end{equation}
Since
\begin{equation}
\frac{\partial L_{p}}{\partial V^{\lambda }}=\sqrt{N}V_{\lambda }
\end{equation}
and
\begin{equation}
\frac{\partial L_{p}}{\partial x^{\lambda }}
=\frac{1}{2} \sqrt{N}g_{\mu \nu,\lambda }V^{\mu }V^{\nu }
+\frac{ 1}{2} \frac{N_{,\lambda }}{N}\, ,
\end{equation}
a straightforward calculation gives the equations of motion of the particle as

\begin{equation}
\frac{d^{2}x^{\mu }}{ds^{2}}+\Gamma _{\nu \lambda }^{\mu }V^{\nu }V^{\lambda
}+\left( V^{\mu }V^{\nu }-g^{\mu \nu }\right) \nabla _{\nu }\ln \sqrt{N}=0
\, .
\end{equation}
As when $\sqrt{N}\rightarrow 1$, (the GR limit), we find the standard geodesic motion.\\

For a barotropic fluid equation of state,
$p=w\rho$  and as $w\ll 1$, so $p+ \rho \approx \rho$ and $T=\rho
-3p\approx \rho$. Moreover, we suppose that the function $f_T$ is a function of $T\approx \rho $
only. We  expand function $f_T$ near a critical point $\rho_0$ :
\begin{eqnarray}
f_T\left(\rho \right)=f_T\left(\rho _0\right)
+\left(\rho -\rho _0\right)f_{TT}
|_{\rho=\rho_0} =2\kappa^2 \left[\alpha_0+ \beta_0\left(\rho -\rho
_0\right)\right]
\end{eqnarray}
here $\alpha_0=f_T\left(\rho _0\right)/2\kappa^2 $ and
$\beta_0=f_{TT}|_{\rho=\rho_0}/2\kappa^2$. With this assumption we obtain:
\begin{equation}
\sqrt{N} \approx c_1 \left[ {{\rho ^{\left( {\frac{{{\rm{ }}w(1 + {a_0} - {b_0}{\rho _0})}}{{1 + 3{a_0} - 3{b_0}{\rho _0}}}} \right)}}{{\left\{ {1 + 3{a_0} - 3{b_0}{\rho _0} + 3{b_0}\rho } \right\}}^{\left( {\frac{{ - 2w}}{{1 + 3{a_0} - 3{b_0}{\rho _0}}}} \right)}}} \right],
\end{equation}
Furthermore, the above equation can be written as
\begin{equation}\label{Q2}
\sqrt N  \approx {c_1}{\rho ^{\left( {\frac{{w(1 + {a_0} - {b_0}{\rho _0})}}{{1 + 3{a_0} - 3{b_0}{\rho _0}}}} \right)}} + O\left( {{\rho ^{ - 2}}} \right),
\end{equation}
Note that the Eq.~(\ref{Q}) is valid in both the Newtonian (non-relativistic) and Modified gravity, GR (the extreme relativistic) regimes.\\

In order to calculate  the Newtonian limit of our $f(\mathcal{R,G,T})$ modified gravity theory. The weak field limit of the gravitational field is well described by the following metric in Newtonian gauge,
\begin{equation}
ds\approx \sqrt{1+2\Phi -\vec{v}^{2}}dt
\approx \left( 1+\Phi -\vec{v}^{2}/2\right) dt\, ,
\end{equation}

here $\Phi $ is the Newtonian potential and $\vec{v}$ stands for the usual velocity (3D) of the fluid.
We can approximate $\sqrt{N}\left(\rho \right)$ given by equation~(\ref{Q2}) as

\begin{eqnarray}
\sqrt N  \approx 1 + \frac{{w\left( {1 + {a_0} - {b_0}{\rho _0}} \right)}}{{1 + 3{a_0} - 3{b_0}{\rho _0}}}\log \left( {{c_1}\rho } \right) = 1 + U(\rho ),
\end{eqnarray}

Here we defined an appropriate Newtonian potential $U(\rho)$. Now we should derive the weak field limit (1st-order) of the equations of motion of a test particle in this force field, it can be explored using
the variational principle

\begin{equation}
\delta \int \left[ 1+U\left(\rho \right)
+\Phi -\frac{\vec{v}^{2}}{2}\right] dt=0\, ,
\end{equation}

By calculating this variational term, we get:

\begin{equation}
\vec{a}=-\nabla \Phi -\nabla U\left(\rho
\right)=\vec{a}_{N}+\vec{a}_p+\vec{a}_{E}\, ,
\end{equation}

here $\vec{a}$ labeled the total net non-relativistic acceleration of the system,and because the gravitational-potential is supposed to be conserved, so we can relate the acceleration to the gradient of the Newtonian potential like $\vec{a}_{N}=-\nabla \Phi $. So we can write the Newtonian gravitational
acceleration as
\begin{equation}
\label{fidr} \vec{a}_p = -\frac{1}{\rho}\nabla{p} ,
\end{equation}
and
\begin{equation}
\vec{a}_{E}\left(\rho ,p\right) = \frac{(b_0 \rho_0 - a_0)}{c_1(1 + 3 a_0 - 3 b_0 \rho_0)}\frac{1}{\rho} \nabla p,
\end{equation}
The extra term is a supplementary acceleration induced due to the modification of the action of the gravitational field.

\subsection{The precession of the perihelion of Mercury}\label{The precession of the perihelion of Mercury}
Solar system tests offer a good test background for the gravitational theories, because the parameters are predictable with a very high accuracy. In our case, the extra-force${\cal{Z}}^{\sigma}$ given in Eq. (\ref{c}) arises by the coupling between matter and geometry. A standard method is to use the invariant properties of the  Laplace-Runge-Lenz vector  (LRL vector),is
defined as

\begin{eqnarray}
\vec{A}=\vec{v}\times \vec{L}-\alpha \vec{e}_{r},
\end{eqnarray}
here by  $\vec{v}$ is the relative velocity from the planet (with mass $m$) to the central Sun (with mass $M_{\odot}$). and the two-body position vector is given by $\vec{r}=r\vec{e}_{r}$ and the unit vector $\vec{e}_{r}=\frac{\vec{r}}{r}$ is the radial unit vector.
As usual, for planetary motion, we use two-body scenario in which the system (Sun-planet) moves with a relative momentum vector $\vec{p}=\mu \vec{v}$ with the "reduced mass" which is defined as $\mu =mM_{\odot }/\left( m+M_{\odot}\right) $. The relative angular momentum $\vec{L}$ is defined in the standard form
 $ \vec{L}=\vec{r}\times \vec{p}=\mu
r^{2}\dot{\theta}\vec{k}$, where $\vec{p}$ is the relative linear momentum of the reduced mass, and $\alpha
=GmM_{\odot}$ \cite{prec}.

With gravitational field, the orbit is an elliptic with
eccentricity $e$, Time period $T$, and major semi-axis $a$. The equation of the
orbit is presented  by $\left( L^{2}/\mu\alpha \right) r^{-1}=1+e\cos
\theta $. The LRL vector can be rewritten as
\begin{equation}
\vec{A}=\left( \frac{\vec{L}^{2}}{\mu r}-\alpha \right)
\vec{e}_{r}-\dot{r}L\vec{e}_{\theta }\, ,
\end{equation}
The derivative of $\vec{A}$ w.r.t the polar angle $\theta $ is related to the effective potential of the central force like
\begin{equation}
\frac{d\vec{A}}{d\theta }
=r^{2}\left[ \frac{dV(r)}{dr}-\frac{\alpha}{r^{2}}\right]
\vec{e}_{\theta
}\, ,
\end{equation}
Where $V(r)$ is potential term and consists of the Post-Newtonian potential, $V_{PN}(r)=-\frac{\alpha}{r}-3\frac{\alpha ^{2}}{mr^{2}},$, plus the additional relativistic contribution from the matter-geometry coupling. This vector quantity is given by:
\begin{eqnarray}
&&\frac{d\vec{A}}{d\theta }=r^2\Big[\frac{6\alpha^2}{mr^3}+m\vec{a}_{E}(\vec{r})\Big]\vec{e}_{\theta}
\end{eqnarray}

The change in direction $\Delta \Phi $ of the perihelion of the planet is written in terms of the a change of $\theta $ from $0$ to  $2\pi $ , like
\begin{eqnarray}
&&\Delta\Phi =\frac{1}{\alpha e}\int_{0}^{2\pi }|\dot{\vec{L}}\times\frac{
d\vec{A}}{d\theta }| d\theta
\end{eqnarray}
By putting the value of $\frac{d\vec{A}}{d\theta }$ and simplifying, we get

\begin{equation}\label{prec}
\Delta \Phi =24\pi ^{3}\left( \frac{a}{T}\right) ^{2}\frac{1}{1-e^{2}}
+\frac{L}{8\pi ^{3}me}
\frac{\left( 1-e^{2}\right) ^{3/2}}{\left( a/T\right)^{3}}
\int_{0}^{2\pi }\frac{a_{E}\left[ L^{2}
\left( 1+e\cos \theta \right)^{-1}/m\alpha \right] }{\left( 1+e\cos \theta
\right)^{2}}
\cos \theta d\theta\, ,
\end{equation}
here we use the identity  $\frac{\alpha}{L}=\frac{2\pi  a}{T
\sqrt{1-e^{2}}}$. As usual, the second term in Eq. (\ref{prec}) gives the contribution to the
perihelion precession through the coupling between
matter-geometry.

Using the Eq.~(\ref{prec}) with the help of Kepler's third law, $T^2=4\pi
^2a^3/GM_{\odot}$. we estimate the perihelion precession:

\begin{equation}\label{prec1}
\Delta \Phi =\frac{6\pi GM_{\odot}}{a\left( 1-e^{2}\right) }
+\frac{2\pi
a^{2}
\sqrt{1-e^{2}}}{GM_{\odot}}a_{E}\, ,
\end{equation}
For the sample planet as Mercury $e=0.205615$, and $a=57.91\times 10^{11}$ cm, respectively, while $M_{\odot }=1.989\times
10^{33}$ g, we estimate the difference $
\left(\Delta \Phi
\right)_{E}=\left(\Delta \Phi \right)_{obs}-\left( \Delta \Phi
\right) _{GR}=0.17(\frac{arcsec}{century})
$
can be attributed to other
physical effects. Hence the observational constraints requires
that the value of the constant $a_E$, must obey the bound as
$a_E\leq 1.28\times 10^{-9} cm/s^2$.

\section{Spherical Anisotropic Fluids and Boundary Conditions}

As we wish to study the influence of anisotropic stress on the mathematical
modeling of the relativistic compact stellar interiors. Having this in mind, we suppose the anisotropic fluid as follow
\begin{align}\label{2l}
{\mathcal{T}_{\sigma \lambda }} = (\rho  + {P_t}){V_\sigma
}{V_\lambda } - {P_t}{g_{\sigma \lambda }} + \Pi{X_\sigma }{X_\lambda },
\end{align}
where $\Pi=(P_r-P_t)$ and $P_r$ is radial pressure while $P_t$ is tangential pressure. The quantities ${V_\sigma}$ and $X_\sigma$
are fluid's four vectors. We let our system in
non tilted frame, which restrict these quantities to satisfy ${V^\sigma }{V_\sigma }=1$ and ${X^\sigma }{X_\sigma }=-1$ relation.
So, the Eq. (\ref{2}) can be rewritten, after assuming $\mathcal{L}_m=\rho$, as
$${\Theta _{\sigma \lambda }} =  - 2{\mathcal{T}_{\sigma \lambda }} + \rho {g_{\sigma \lambda }}$$
With the help of this, the equation of motion (\ref{5}) can be written as
\begin{equation}\label{fieldequation}
{\mathcal{R}_{\sigma \lambda }} - \frac{1}{2}\mathcal{R}{g_{\sigma \lambda }} = \mathcal{T}_{\sigma \lambda
}^{\textit{eff}},
\end{equation}
Here, $\mathcal{T}_{\sigma \lambda}^{\textit{eff}}$ represent the effective energy momentum tensor and is written as
\begin{align}\nonumber
&\mathcal{T}_{\sigma \lambda }^{\textit{eff}} =\frac{1}{{{f_{\cal R}}\left( {{\cal R},{\cal G},{\cal T}} \right)}}\left[ {{\kappa ^2}{{\cal T}_{\sigma \lambda }}} \right. - \left( {{{\cal T}_{\sigma \lambda }} + {\rho g _{\sigma \lambda }}} \right){f_T}\left( {{\cal R},{\cal G},{\cal T}} \right) \\\nonumber
& + \frac{1}{2}{g_{\sigma \lambda }}[f\left( {{\cal R},{\cal G},{\cal T}} \right) + \mathcal{R}{f_{\cal R}}\left( {{\cal R},{\cal G},{\cal T}} \right)] + {\nabla _\sigma }{\nabla _\lambda }{f_{\cal R}}\left( {{\cal R},{\cal G},{\cal T}} \right)\\\nonumber
& - {g_{\sigma \lambda }}\Box{f_{\cal R}}\left( {{\cal R},{\cal G},{\cal T}} \right) - (2\mathcal{R}{\mathcal{R}_{\sigma \lambda }} - 4\mathcal{R}_\sigma ^\xi {\mathcal{R}_{\xi \lambda }} - 4{\mathcal{R}_{\sigma \xi \lambda \eta }}{\mathcal{R}^{\xi \eta }}\\\nonumber
& + 2\mathcal{R}_\sigma ^{\xi \eta \delta }{R_{\lambda \xi \eta \delta }}){f_{\cal G}}\left( {{\cal R},{\cal G},{\cal T}} \right) - (2\mathcal{R}{g_{\sigma \lambda }}{\nabla ^2} - 2\mathcal{R}{\nabla _\sigma }{\nabla _\lambda } - 4{g_{\sigma \lambda }}{\mathcal{R}^{\xi \eta }}{\nabla _\xi }{\nabla _\eta }\\
& - 4{\mathcal{R}_{\sigma \lambda }}{\nabla ^2} + 4\mathcal{R}_\sigma ^\xi {\nabla _\lambda }{\nabla _\xi } + 4\mathcal{R}_\lambda ^\xi {\nabla _\sigma }{\nabla _\xi } + 4{\mathcal{R}_{\sigma \xi \lambda \eta }}{\nabla ^\xi }{\nabla ^\eta })\left. {{f_{\cal G}}\left( {{\cal R},{\cal G},{\cal T}} \right)} \right],
\end{align}

Furthermore, we consider the static spherical symmetric geometry with the line-element written as
\begin{equation}\label{zz7}
d{s^2} = {e^{a_1}}d{t^2} - {e^{a_2}}d{r^2} - {r^2}\left[ {d{\theta ^2} + {{\sin }^2}\theta d{\phi ^2}} \right],
\end{equation}
where $a_1$ and $a_2$ are radial dependent functions. We assume that this relativistic geometry (Eq. \ref{zz7}) with the anisotropic matter distribution. The functions $a_1$ and $a_2$ can be further rewritten as more specific combinations of their arguments. Moreover, Ref. \cite{zs43} proposed the mathematical formulations of these two variables in terms of three constants $A_1,~A_2$ and $A_3$ as $a_1 =A_2 r^2 +A_3$ and $a_2=A_1 r^2$. The values of these parameter constants can be calculated by taking some observational values of compact stellar configurations.\\

In order to design our system, such that, it is defined with a 3D time-like boundary surface, $\Im$ and the interior-geometry to $\Im$ is denoted with the help of an Eq.(\ref{zz7}), while the exterior-geometry to this hyper-surface $\Im$ is represented with Schwarzschild vacuum solution, like
\begin{equation}\label{ex1}
d{S^2} = \left[ {1 - \frac{{2M}}{r}} \right]d{t^2} - \frac{d{r^2}}{{\left[ {1 - \frac{{2M}}{r}} \right]}} - {r^2}\left[ {d{\theta ^2} + {\mathop{\rm \sin}\nolimits} {\theta ^2}d{\varphi ^2}} \right]
\end{equation}
Here, $M$ represents mass of black-hole. As we have two geometries, interior and exterior and we need to joint them at specific boundary surface. So, using the continuity of metric coefficients $g_{tt},~g_{rr}$ and its derivative $\frac{\partial g_{tt}}{\partial r}$ over $\Im$. The continuous joining of interior and exterior geometry over $\Im$ at $r=R$ gives the following constraints
\begin{align}
A_1 &= -\frac{{1}}{{{R^2}}}\ln \left[ {1 - \frac{{2M}}{R}} \right], \quad
A_2 = \frac{M}{{{R^3}}}{\left[ {1 - \frac{{2M}}{R}} \right]^{ - 1}},\\
A_3 &= \ln \left[ {1 - \frac{{2M}}{R}} \right] + \frac{M}{R}{\left[ {\frac{{2M}}{R}-1} \right]^{ - 1}}.
\end{align}
We accomplish our analysis by using up the values of constants $A_1,~A_2$ and $A_3$ with numerous systematic grounds. For this purpose, we assume three compact stars candidates, i) $Her X-1$, ii) $SAXJ1808.4-3658$ and iii) $4U1820-30$. We see that all these relativistic compact star candidates have the ratio $\frac{2M}{R}$ less than $\frac{8}{9}$. With the help of these observational data \cite{dey2013,guver2010b}, we calculate various structural values to analyze different realistic aspects of stellar structures.

\begin{table}[h!]
\centering
\begin{tabular}{|c| c| c| c| c| c|}
 \hline
CS  &$M$ & $R$ & $\mu=\frac{M}{R}$ &$A_1$ &$A_2$ \\ [0.5ex]
\hline\hline\
Her X-1  & $0.88$ & $7.7$ & $0.168$ & $0.006906276428 $ &$0.00426736461 $\\ [1ex]
\hline\
SAXJ1808.4-3658& $1.435$ & $7.07$ & $0.299$ & $0.01823156974$ &$0.0148801156$\\ [1ex]
\hline
4U1820-30& $2.25$ & $10$ & $0.332$ & $0.01090644119 $ &$0.00988095238$\\ [1ex]
\hline
\end{tabular}
\caption{Compact stars $(CS)$ and their observed values of the masses in $(M_{\odot})$, radii in $km$, compactness $\mu$, and the constants $A_1$, and $A_2$ in $km^{-2}$.}
\label{table:1}
\end{table}

\subsection{Different Models}

In order to investigate some cosmological aspects of $f(\mathcal{R,G,T})$ gravity  along with the better understanding in mathematical modeling of compact stars, we need to take viable model in this theory. The consequences from our outcomes may lead us to reveal the several unseen cosmological results at both astrophysical scales and theoretical, depends upon the choice of $f(\mathcal{R,G,T})$ models.\\
For this purpose, we shall choose three different models in $f(\mathcal{R,G,T})$ gravity, like
\begin{equation}
f(\mathcal{R,G,T})=f(\mathcal{R})+f(\mathcal{G})+\lambda\mathcal{T}
\end{equation}
In Ref. \cite{new1}, choosing $f(\mathcal{R})=\mathcal{R} + \alpha \mathcal{R}^2$ in such a way provides the leverage that analyze feasibility of compact stars for different values of $\alpha$ by avoiding intermediate approximation in system of equations. It is exhibited through analysis that for $\alpha>0$, the scalar curvature decreases exponentially with respect to distance and hence does not decreases to zero at the surface of star, which is a good agreement with general theory of relativity. Additionally, with increase in $\alpha$ the stellar mass confined by surface of the star decreases. Moreover, keeping in view the Einstein predictions, decrease in the mass confined by surface of compact stars may leads to decrement in surface redshift in $R^2$- gravity background. Also, straight choice of $f(R)$ gravity enables to analyze corresponding matching conditions at edge of compact stars and does not require vanishing of the scalar curvature.
\begin{itemize}
  \item \textbf{Model-(a)}\\
Firstly, we assume the following model in $f(\mathcal{R,G,T})$  gravity
\begin{equation}\label{model1}
f(\mathcal{R,G,T})=\mathcal{R} + \alpha \mathcal{R}^2 + \beta \mathcal{G}^n + \gamma \mathcal{G} \ln(\mathcal{G})+\lambda \mathcal{T},
\end{equation}
In which $\alpha,\beta,\gamma,\lambda$ and $n$ are the constant parameters of model.\\
In this model we are taking Gauss-Bonnet curvature scalar $\mathcal{G}^{n}$, with $n$ is the real constant and possesses a field equation which is scale-invariant. This yields to be one of the possible analogies to $f(\mathcal{R})$-theories. In this suggested model $\mathcal{G}^{n}$ is multiplied with a constant $\beta$. Additionally, the logarithmic term should be dimensionless, so alternatively we should have taken the logarithmic term as $log(\frac{\mathcal{G}}{\mathcal{G}_{0}})$. The change of the value $\mathcal{G}_{0}$ can be avoided by redefining the constant $\beta$ because the term with $\beta$ does not play role in the field equation. The term with constant $\beta$ can be allowed to set as zero classically. \\
  \item \textbf{Model-(b)}\\
  Secondly, we take the following model in $f(\mathcal{R,G,T})$  gravity
\begin{equation}\label{model2}
f(\mathcal{R,G,T})= \mathcal{R} + \alpha \mathcal{R}^2 + \beta \mathcal{G}^{n_2} (1 + \gamma \mathcal{G}^{m_2})+\lambda \mathcal{T},
\end{equation}
In which $\alpha,\beta,\gamma,\lambda$, $n_2$ and $m_2$ are the constant parameters of model.\\
It is indeed provocative to find the conditions on $n_2,m_2, \beta$ and $\gamma$ in order to avoid singularities of any type. The model can be written as $f(\mathcal{R,G,T})\sim \mathcal{R} + \alpha \mathcal{R}^2 +\beta G^{n_{3}}+\lambda \mathcal{T}$ for $G\rightarrow \pm \infty$ or $G\rightarrow 0^{-}$. It is mandatory to mention here that we do not consider the trivial case $n_2=m_2$, and taking $n_2>0$. \\
  \item \textbf{Model-(c)}\\
  Similarly, we let another viable model in $f(\mathcal{R,G,T})$  gravity
\begin{equation}\label{model3}
f(\mathcal{R,G,T})= \mathcal{R} + \alpha \mathcal{R}^2 + (a_1 \mathcal{G}^{n_3} + b_1)/(a_2 \mathcal{G}^{n_3} + b_2)+\lambda \mathcal{T}.
\end{equation}
In which $\alpha,\beta,\lambda$, $a_1$, $a_2$, $b_1$, $b_2$ and $n_3$ are the constant parameters of model.
For large values of $\mathcal{G}$, it gravitates to a constant and hence to find out singularities becomes impossible (it is established that $R$+constant is singularities free, pursuant to $\Lambda$CDM model). It is still possible that singularities can occur when $G\rightarrow 0^{-}$ for $n_3>1$.
\end{itemize}
The last terms $\lambda\mathcal{T}$ is a linear term and some of the physical properties of the stars, such as pressure, energy density, mass and radius, are affected when $\lambda$ is changed. While for a fixed central star energy density, the mass of neutron (and strange stars) can increase with $\lambda$. Concerning the star radius, it increases for neutron stars and it decreases for strange stars with the increment of $\lambda$. This specific choice of models avoids tedious calculations and consequently the results reveal several unknown cosmological phenomenon on both astrophysical and theoretical scales according to choice of aforesaid terms. Moreover the suggested models have the tendency to describe both the early universe and late universe.\\

\section{Physical analysis}

Here, in this section, we investigate various aspects of relativistic compact stellar structures. For this, we let three distinct structures, named as $Her X-1$, $SAX J 1808.4-3658$, and $4U 1820-30$ with their masses as $0.88M_{\odot},~1.435M_{\odot}$ and $2.25M_{\odot}$, respectively.\\
Along with this data, we use three different models $(a)$, $(b)$ and $(c)$ and evaluate the field equation which yields the matter variables in the term of model parameters.\\
Further, we study different physical features to get more realistic configurations of these stellar structures. The outcomes from these modeling may give the evidence for the viability of the models in $f(\mathcal{R,G,T})$ gravity on astrophysical as well as on theoretical grounds.\\
For simplicity, the relativistic compact stellar structures $Her X-1$, $SAX J 1808.4-3658$, and $4U 1820-30$ are labeled as \emph{CS-I}, \emph{CS-II} and \emph{CS-III}, respectively.

\subsection{Density and stresses profile}

\begin{figure} \centering
\epsfig{file=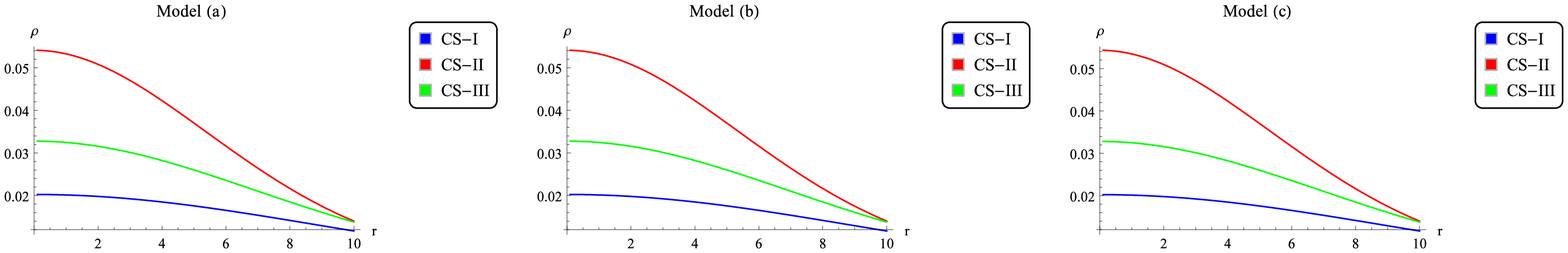,width=1\linewidth}
\caption{The energy density of the strange star candidates \emph{CS-I}, \emph{CS-II} and \emph{CS-III}.}\label{roc}
\epsfig{file=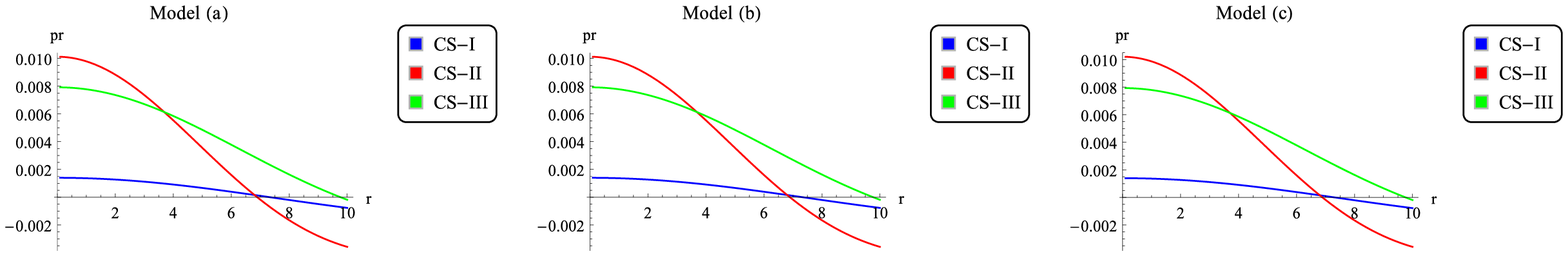,width=1\linewidth}
\caption{The radial pressure of the strange star candidates \emph{CS-I}, \emph{CS-II} and \emph{CS-III}.}\label{prc}
\epsfig{file=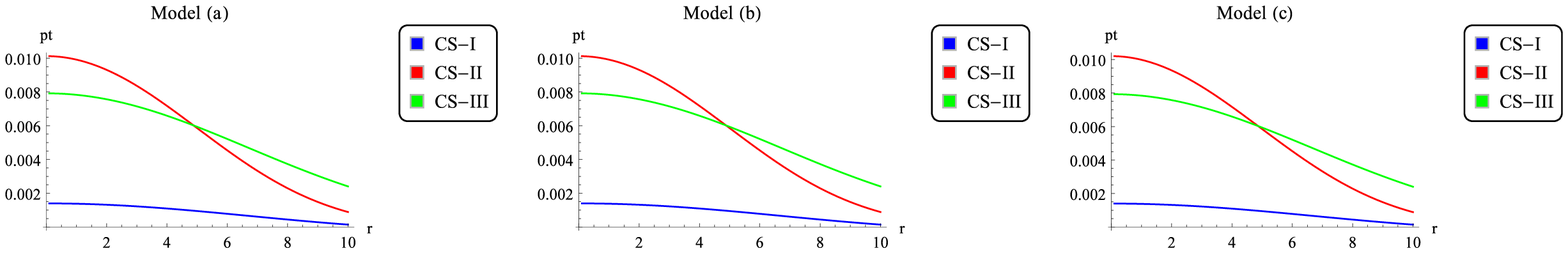,width=1\linewidth}
\caption{The transverse pressure of the strange star candidates \emph{CS-I}, \emph{CS-II} and \emph{CS-III}.}\label{ptc}
\end{figure}

\begin{figure} \centering
\epsfig{file=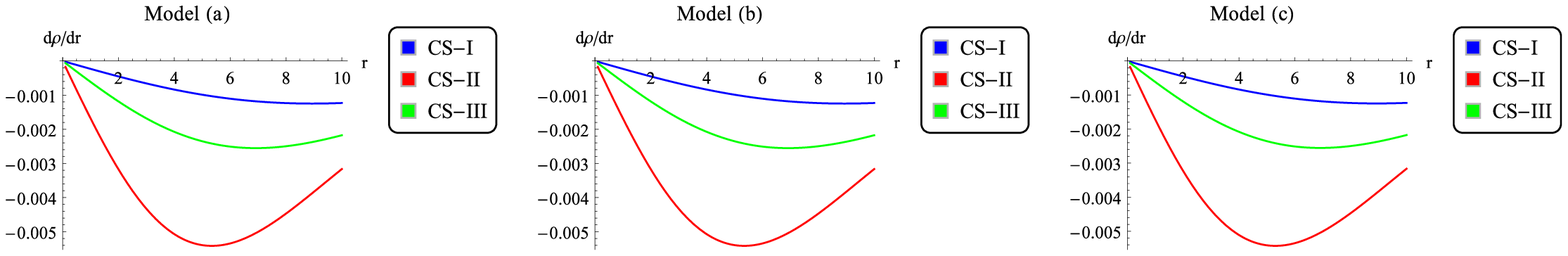,width=1\linewidth}
\caption{The behavior of $d\rho/dr$ of the strange star candidates \emph{CS-I}, \emph{CS-II} and \emph{CS-III}.}\label{dro}
\epsfig{file=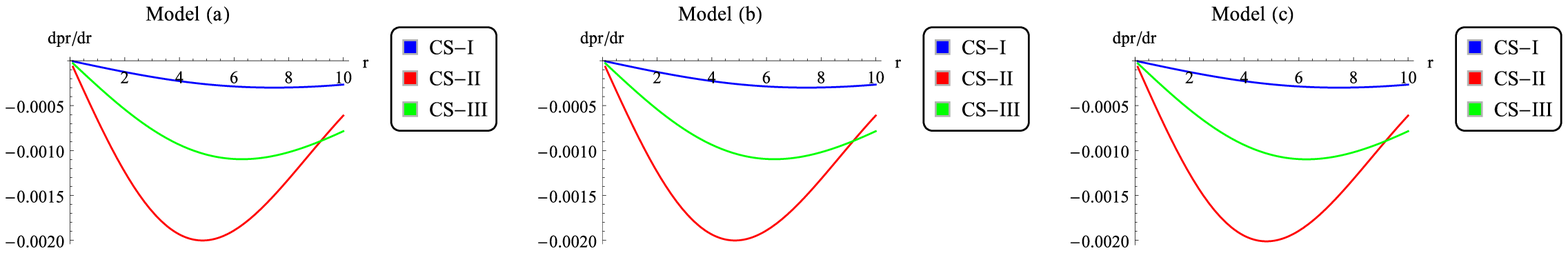,width=1\linewidth}
\caption{The behavior of $dP_r/dr$ of the strange star candidates \emph{CS-I}, \emph{CS-II} and \emph{CS-III}.}\label{dpr}
\epsfig{file=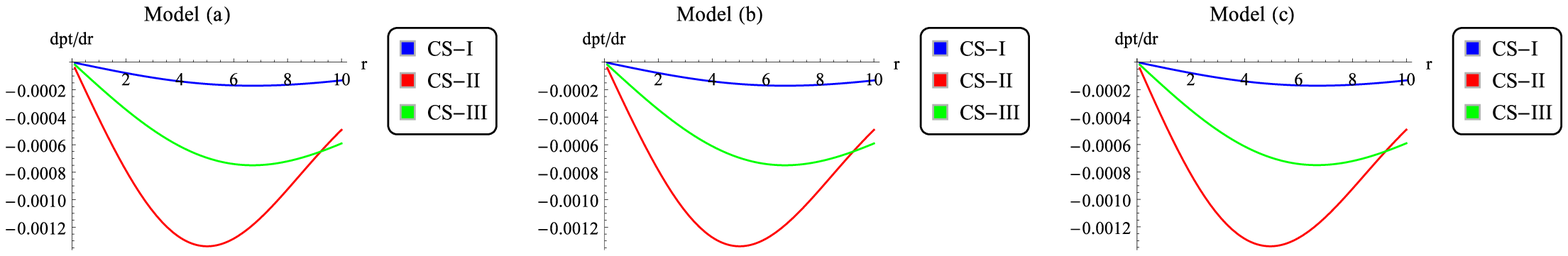,width=1\linewidth}
\caption{The behavior of $dP_t/dr$ of the strange star candidates \emph{CS-I}, \emph{CS-II} and \emph{CS-III}.}\label{dpt}
\end{figure}

Here, we study the behavior of different structural variables of \emph{CS-I}, \emph{CS-II} and \emph{CS-III} which varies with the radial coordinate. We shown through different plots that the evolutions are not only in the density and anisotropic stresses but also in their radial derivative. We plot these variations for \emph{CS-I}, \emph{CS-II} and \emph{CS-III} with \emph{Model-$(a)$}, $(b)$ and $(c)$.\\
As in fig. (\ref{roc}), we see that the energy density of \emph{CS-I}, \emph{CS-II} and \emph{CS-III} are increasing with the decreasing of radial coordinate. The density in each case, becomes so maximum at the core of these stars which make them more compact and denser. The density is monotonically decreasing function with the increase in radial coordinate $r$, as a result, the density at the surface of these stars become less dense than the core.  This validate the high compactness of \emph{CS-I}, \emph{CS-II} and \emph{CS-III} cores, thus validate all of our models in $f(\mathcal{R,G,T})$ gravity.\\
The variation in the radial pressure and traverser stellar pressure are shown in fig. (\ref{prc}) and (\ref{ptc}).
By the similar fashion, the plots shown in Figures (\ref{dro}), (\ref{dpr}) and (\ref{dpt}) investigate the behavior of the variation of radial derivative of density and anisotropic pressures. We can see that $\frac{d\rho}{{dr}}<0$, $\frac{dP_r}{{dr}}<0$ and $\frac{dP_t}{{dr}}<0$ for all model $(a)$, $(b)$ and $(c)$ and \emph{CS-I}, \emph{CS-II} and \emph{CS-III}. Furthermore, the radial derivative vanishes at $r=0$ e.g.
$$\frac{d\rho}{{dr}}=0, \frac{dP_r}{{dr}}=0.$$
Moreover, 2nd derivative of all these variables yields negative results. These results advocate the hefty profiles of stellar matter variables, so representing compact environments of \emph{CS-I}, \emph{CS-II} and \emph{CS-III}.

\subsection{Energy Conditions}

\begin{figure}
\centering
\epsfig{file=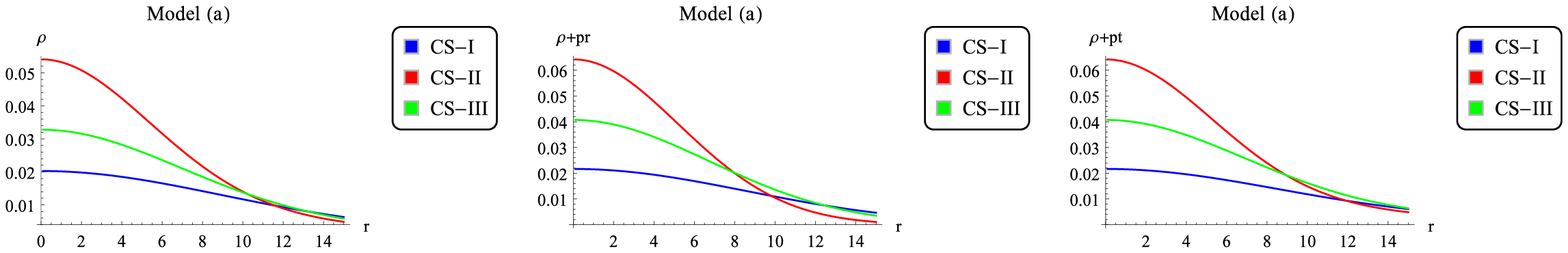,width=1\linewidth}
\epsfig{file=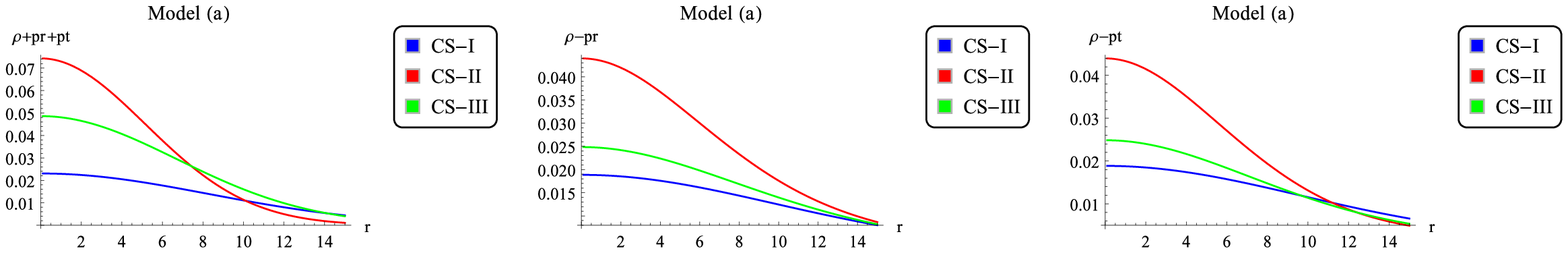,width=1\linewidth}
\caption{Different energy conditions for \emph{Model-$(a)$}.}\label{energy1}
\end{figure}

\begin{figure}
\centering
\epsfig{file=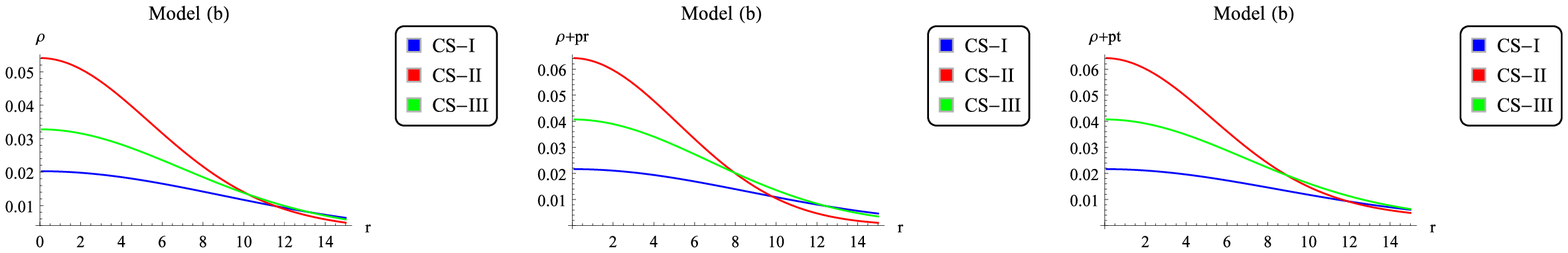,width=1\linewidth}
\epsfig{file=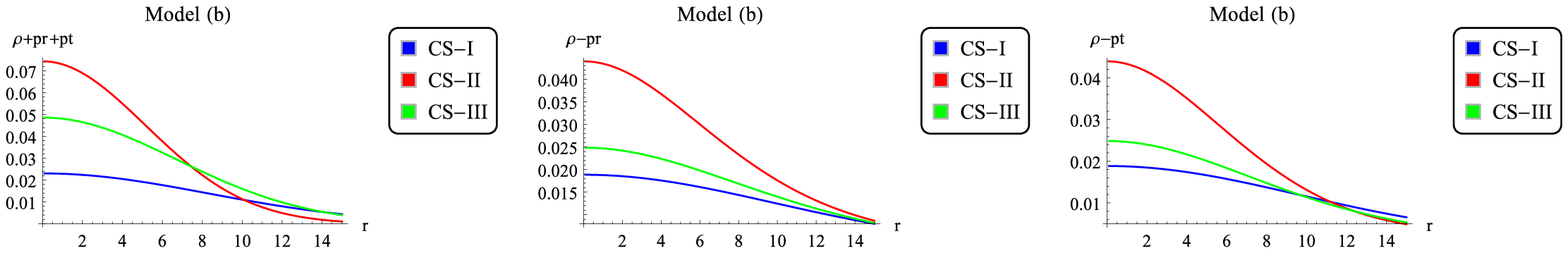,width=1\linewidth}
\caption{Different energy conditions for \emph{Model-$(b)$}.}\label{energy2}
\end{figure}

\begin{figure}
\centering
\epsfig{file=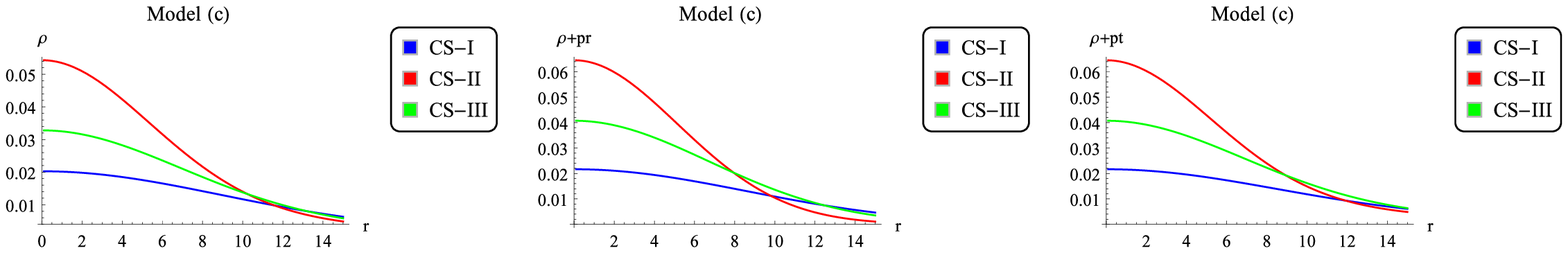,width=1\linewidth}
\epsfig{file=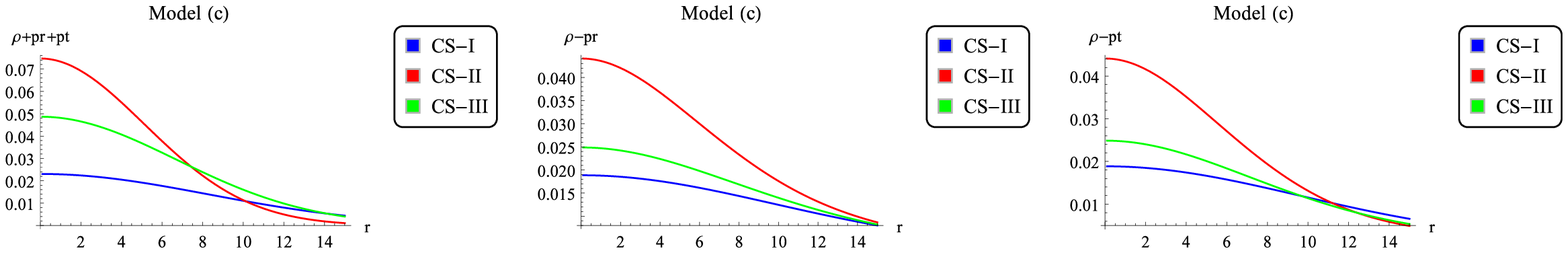,width=1\linewidth}
\caption{Different energy conditions for \emph{Model-$(c)$}.}\label{energy3}
\end{figure}

In order to check the matter content distribution is realistic or not, there are some mathematical conditions which imposed on stress energy momentum tensor. These mathematical conditions are widely known as energy conditions.
In the context of $f(\mathcal{R,G,T})$ gravity, these energy conditions are:
\begin{align}\nonumber
&\textrm{{Null Energy Condition: (NEC)}}\Rightarrow{\rho} + {P_i} \ge 0,\\\nonumber
&\textrm{{Weak Energy Condition: (WEC)}}\Rightarrow {\rho} \ge 0 \text{ and } {\rho} +{P_i} \ge 0,\\\nonumber
&\textrm{{Dominant Energy Condition: (DEC)}} \Rightarrow{\rho} \ge 0 \text{ and } {\rho} \pm{P_i} \ge 0,\\\nonumber
&\textrm{{Strong Energy Condition: (SEC)}}\Rightarrow {\rho+2{P_t}+3{P_r}} \ge 0 \text{ and } {\rho} +{P_i} \ge 0.
\end{align}
In our case, we see from Figures (\ref{energy1}), (\ref{energy2}) and (\ref{energy3}) that all the energy conditions are well satisfied under model $(a)$, $(b)$ and $(c)$ for \emph{CS-I}, \emph{CS-II} and \emph{CS-III}. It is to be noted that the anisotropic fluid as in Eq.(\ref{2l}) represent the realistic ground for gravitational sources.


\subsection{TOV Equation}

\begin{figure} \centering
\epsfig{file=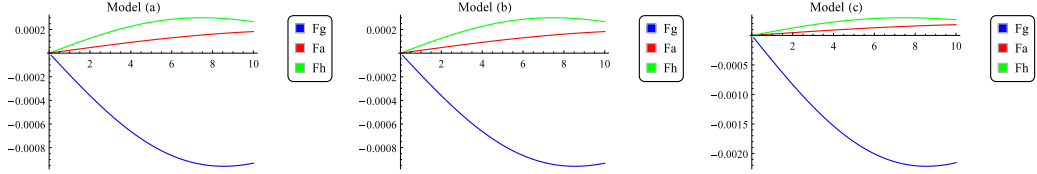,width=1\linewidth}
\caption{The behavior of different forces.}\label{eqb}
\end{figure}

Here, we investigate the equilibrium scenario of compact star geometries. The equilibrium picture for these structures can be achieved by solving Tolman-Oppenheimer-Volkov equations widely known as TOV equation, found as
\begin{equation}\label{tovh}
\frac{{d{P_r}}}{{dr}} + \frac{{{a_1}'( {P_r}+ \rho)}}{2} - \frac{{2({P_t} - {P_r})}}{r} = 0,
\end{equation}
It is to be noted that the TOV equation (\ref{tovh}) contain the role of three well-known forces, e.g. hydrostatic, $F_h$, gravitational, $F_g$,  and anisotropic, $F_a$, forces. Hence, equation (\ref{tovh}) can be rewritten, in general, as
\begin{equation}
F_h + F_g + F_a= 0,
\end{equation}
By the similar fashion, we plot the graphs for \emph{CS-I}, \emph{CS-II} and \emph{CS-III} as shown in Figure (\ref{eqb}).
In the modeling of stellar structures, the equilibrium picture is very important. One can see the variations in these forces over radial coordinate $r$. In Figure (\ref{eqb}), the left is for model $(a)$, middle one is for model $(b)$ while the right plots is for model $(c)$.

\subsection{Stability Analysis}
The relativistic compact structures are more significant which are stable against the variations. So, the stability is the key role in finding the realistic model of stellar structures.
Here, we check the stability of \emph{CS-I}, \emph{CS-II} and \emph{CS-III} by adopting the techniques based on the notion of cracking or overturning procedure. According to this, the two variables radial sound speeds, $v^2_{sr}$, and transverse sound speeds, $v^2_{st}$, need to be in closed interval [0,1].  These two variables are defined as
$$\frac{{d{P_r}}}{{d\rho }} = v_{sr}^2$$
and
$$\frac{{d{P_t}}}{{d\rho }} = v_{st}^2.$$
For the stable configuration, the radial sound speeds as well as the transverse sound speeds must obey $0 \le v_{sr}^2 \le 1$ and $0 \le v_{st}^2 \le 1$ limits.
\begin{figure} \centering
\epsfig{file=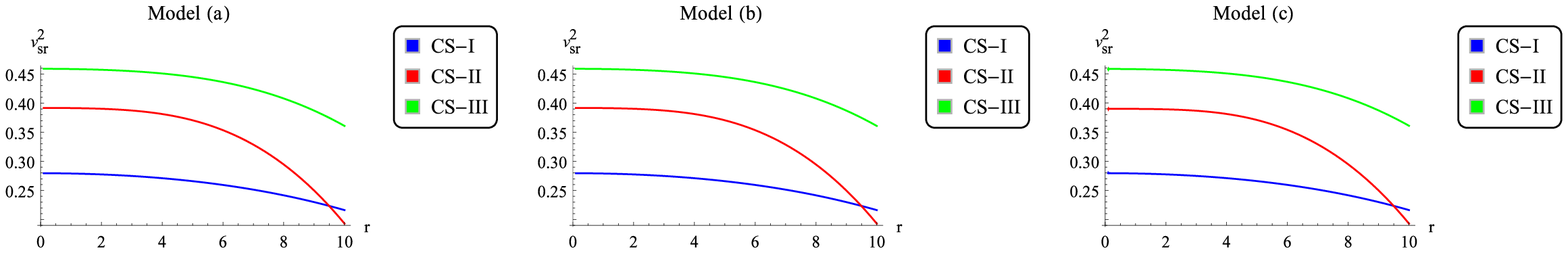,width=1\linewidth}
\caption{Evolutions in $v_{sr}^2$ with respect radius $r$.}\label{vsr}
\end{figure}
\begin{figure} \centering
\epsfig{file=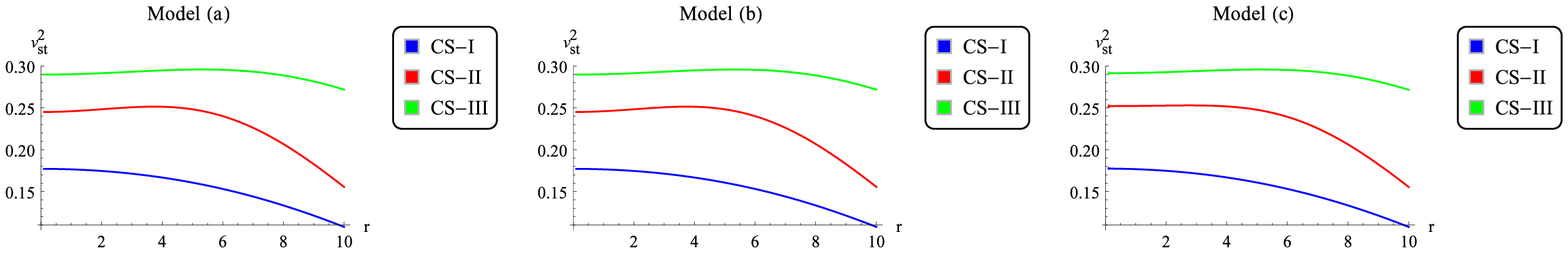,width=1\linewidth}
\caption{Evolutions in $v_{st}^2$ with respect radius $r$.}\label{vst}
\end{figure}
\begin{figure} \centering
\epsfig{file=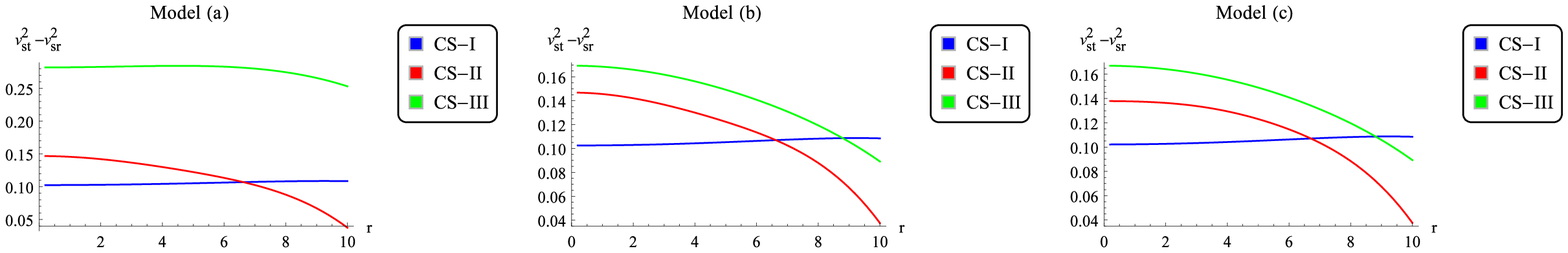,width=1\linewidth}
\caption{Evolutions in of $v_{st}^2 - v_{sr}^2$ with respect radius $r$.}\label{vstmvsr}
\end{figure}
The results are plotted, we see that from the Fig. (\ref{vsr}) and (\ref{vst}) that $v_{sr}^2 $ and $v_{st}^2 $ are within the stability limit for \emph{CS-I}, \emph{CS-II} and \emph{CS-III} stellar structures.
Furthermore, the stability modes are shown in Figure (\ref{vstmvsr}) as follow
$$0< |v_{st}^2-v_{sr}^2|<1.$$
In the context of $f(\mathcal{R,G,T})$ gravity models, this confirms that these relativistic stellar structures are completely in the desire range of stability.

\subsection{EoS Parameter}
\begin{figure} \centering
\epsfig{file=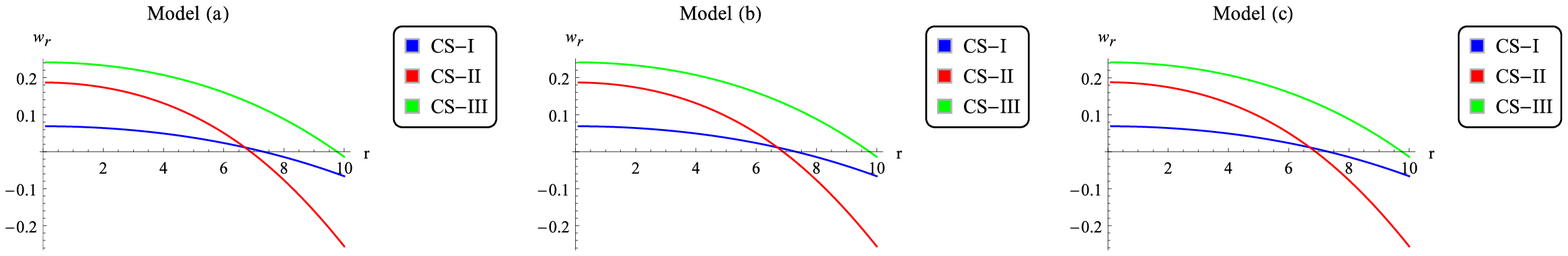,width=1\linewidth}
\caption{Radial EoS parameter.}\label{wr}\end{figure}

\begin{figure} \centering
\epsfig{file=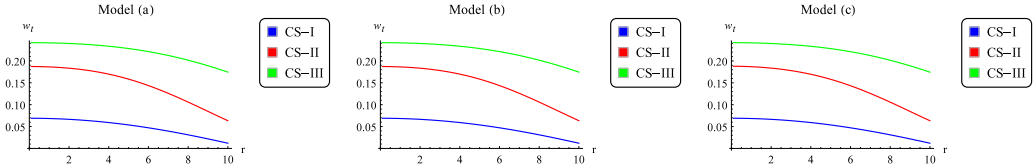,width=1\linewidth}
\caption{Transverse EoS parameter.}\label{wt}\end{figure}

Now for anisotropic case, the equation of state (EoS) are found as
$${w_r} = \frac{{{p_r}}}{\rho },\\
{w_t} = \frac{{{p_t}}}{\rho }.$$
For which the limits is like $0<w_r<1$ and $0<w_t<1$. The behavior of $w_r$ and $w_t$ are shown graphically in figure \ref{wr} and \ref{wt}.
which are in bound and we can safely says that the matter inside these stars are usual charged matter.


\subsection{ Mass Radius Relationship}
\begin{figure} \centering
\epsfig{file=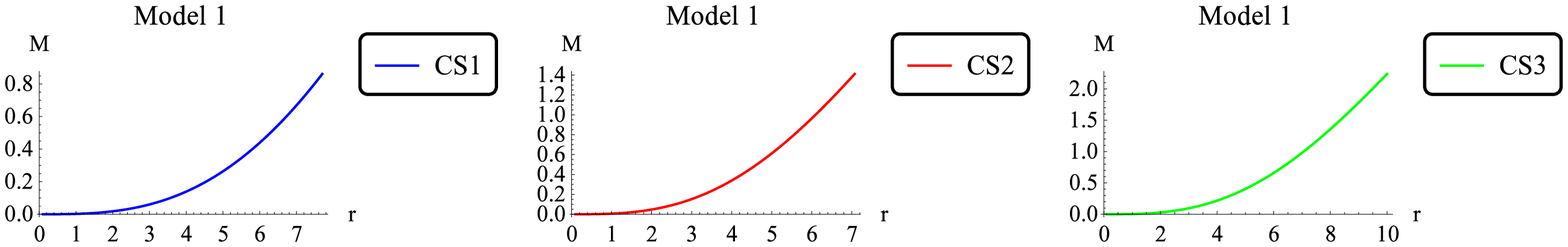,width=1\linewidth}
\epsfig{file=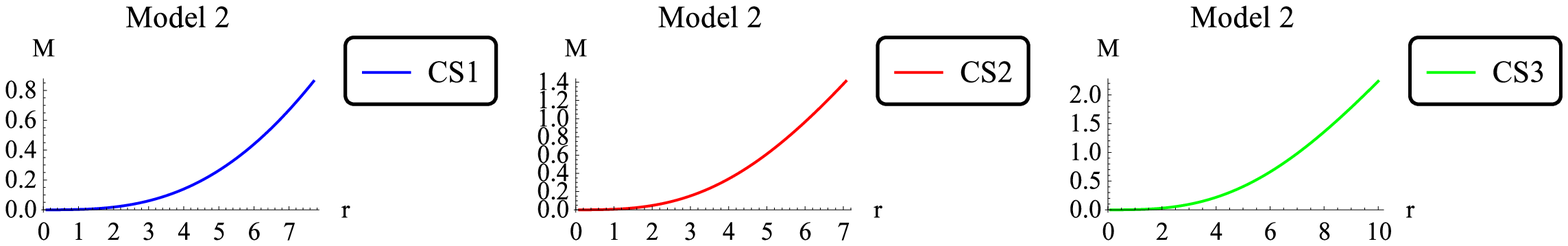,width=1\linewidth}
\epsfig{file=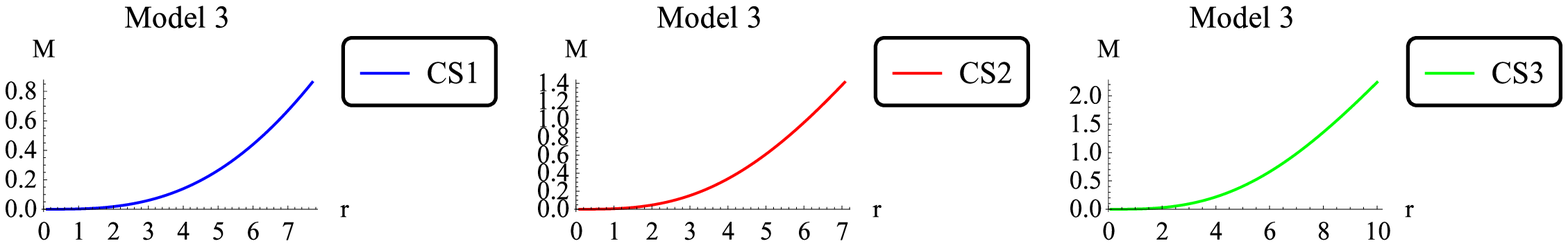,width=1\linewidth}
\caption{Variations of the mass function for different compact stars.}\label{mass}
\end{figure}

The mass of compact stars can be written as
\begin{equation}
m(r) = \int\limits_0^r {4\pi {{r'}^2}\rho dr'}
\end{equation}
Here we know that mass $m$ is function of $r$ and $m(r=0)=0$ while $m(r=R)=M$. The variation in masses of compact stars are shown in Fig. \ref{mass}.
We see that the mass is regular at core because it is directly proportional to radial distance e.g.  $m(r)\rightarrow0$ for $r\rightarrow0$. The maximum mass is attained at $r=R$, as shown in fig. \ref{mass}.

\subsection{Anisotropicity}

\begin{figure} \centering
\epsfig{file=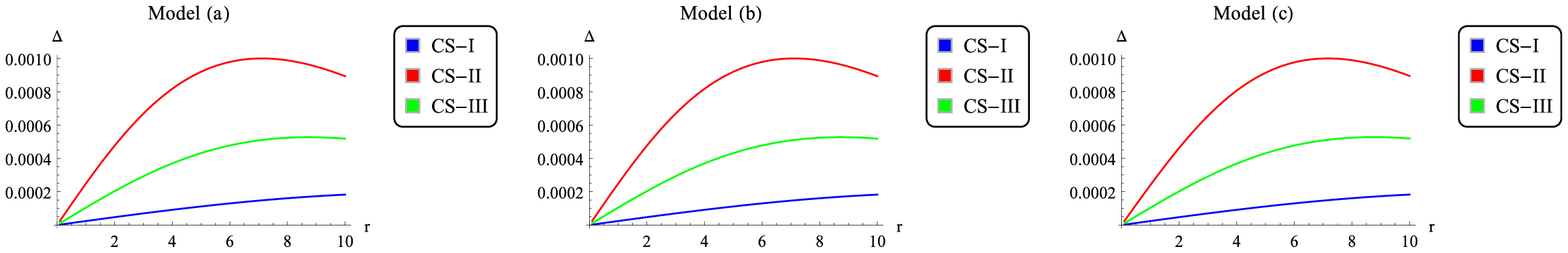,width=1\linewidth}
\caption{Behavior of anisotropic measure $\Delta$.}\label{ani}
\end{figure}

For the mathematical modeling of relativistic stellar structures, the measure of an anisotropy is very important. In the local anisotropic fluid distributions, the magnitude as well as degree of anisotropicity, $\Delta $, can be defined as
\begin{align}\nonumber
\Delta  &\equiv \frac{2}{r}({P_t} - {P_r}).
\end{align}
In our case, we see that $\Delta$ is remained positive for all \emph{CS-I}, \emph{CS-II} and \emph{CS-III} as shown in Figure (\ref{ani}).
As the positive measure of anisotropy implies that the role of transverse pressure is greater than that of radial pressure and hence directed outward.

\section{Final Remarks}

In this work, with an arbitrary coupling between geometry and matter, we offered a new and more generalize gravitational theory $f(\mathcal{R,G,T})$. For this, the Lagrangian is written with the help of modified Hilbert- Einstein action, in which $\mathcal{R}$ is replaced by the arbitrary function $f(\mathcal{R,G,T})$. Furthermore, we formulated the field equation with the help of variational principle and also get the non-zero covariant divergence of stress energy momentum tensor, $\mathcal{T}_{\alpha\beta}$ , which is also consistent with modified $f(R,T)$ gravity \cite{17}.
Moreover, we assume different compact geometries and check the impact of different models in this theory. For this purpose, we consider three distinct relativistic compact stars, named as $Her X-1$, $SAX J 1808.4-3658$, and $4U 1820-30$ and check their physical features. Some of the main results are: \\
We get the result that the energy density is maximum at the core of these stars while minimum at the surface. Similarly, the anisotropic stresses, $P_r, P_t$, also minimum at the surfaces and become maximum at the core. The radial pressure $P_r$ is vanished at the surface of these stars. Due to the very high density ($\rho_c$), at core, these stars become more compact and this validate our models in $f(\mathcal{R,G,T})$.\\
In this study, we checked the impact of extra force which arises from the high curvature and get the matter content more realistic, as in case of all these stars, the energy conditions e.g. \emph{NEC, WEC, DEC, SEC} were well satisfied which validate that the matter destitution is normal matter (not any kind of exotic matter). The significance of energy conditions in the regime of MTG were studied in Ref \cite{ref48, ref49} while in case of Non-local gravity were investigated in ref \cite{ref50}.\\
The stability analysis is very important for the modeling of any compact geometry. For stability, we followed the "cracking techniques", we confirmed that our model is potentially stable against the fluctuations. \\
There are three different kind of forces arise which balance each other, the repulsive anisotropic force $F_a$ acts along the outward direction to balance the joint effect of gravitational force $F_g$ and the hydrodynamic force $F_h$ acting along the inward direction. These forces combine to attain equilibrium.\\
In case of anisotropy, there are two types of equation of state parameters e.g. the radial, $w_r$ and the transverse equation of state parameter $w_t$. Both these equation of state parameters lies in limit of realistic normal matter distribution.\\
The overall remark is that our proposed models in $f(\mathcal{R,G,T})$ gravity satisfy all physical aspects. The whole analysis has been made in connection to direct comparison of various of the compact star candidates, which confirms validity of our proposed model.
Our results are compatible with the results obtained in case of different MTGs \cite{ref19b,refmia1,refmia2,mianew}.  Hence our approach leads to a better analytical description of the compact stars.

\section{Appendix A}

By solving the field equation written in Eq. (\ref{fieldequation}) for the metric Eq. (\ref{zz7}), we get
\begin{align}\nonumber
\rho &= \frac{1}{{2{r^2}}}{{\rm{e}}^{ - 2{a_2}}}( - {{\rm{e}}^{2{a_2}}}f{r^2} + {{\rm{e}}^{2{a_2}}}{r^2}{f_\mathcal{G}}\mathcal{G} + {{\rm{e}}^{{a_2}}}{f_\mathcal{R}}({{\rm{e}}^{{a_2}}}{r^2}\mathcal{R} + 2( - 1 + {{\rm{e}}^{{a_2}}}\\\nonumber
&+ r{{a'}_2})) + 12{{a'}_2}{f_\mathcal{G}}^\prime  - 4{{\rm{e}}^{{a_2}}}{{a'}_2}{f_\mathcal{G}}^\prime  - 4{{\rm{e}}^{{a_2}}}r{f_\mathcal{R}}^\prime  + {{\rm{e}}^{{a_2}}}{r^2}{{a'}_2}{f_\mathcal{R}}^\prime \\\nonumber
&- 8{f_\mathcal{G}}^{\prime \prime } + 8{{\rm{e}}^{{a_2}}}{f_\mathcal{G}}^{\prime \prime } - 2{{\rm{e}}^{{a_2}}}{r^2}{f_\mathcal{R}}^{\prime \prime }),\\\nonumber
P_r &= \frac{1}{{2\left( {1 + {f_\mathcal{T}}} \right){r^2}}}{{\rm{e}}^{ - 2{a_2}}}({{\rm{e}}^{2{a_2}}}f{r^2} + {{\rm{e}}^{2{a_2}}}f{f_\mathcal{T}}{r^2} - {{\rm{e}}^{2{a_2}}}(1 + {f_\mathcal{T}}){r^2}{f_\mathcal{G}}\mathcal{G}\\\nonumber
&- {{\rm{e}}^{{a_2}}}{f_\mathcal{R}}({{\rm{e}}^{{a_2}}}\left( {1 + {f_\mathcal{T}}} \right){r^2}\mathcal{R} + 2(( - 1 + {{\rm{e}}^{{a_2}}})\left( {1 + {f_\mathcal{T}}} \right) - r{{a'}_1}\\\nonumber
&+ {f_\mathcal{T}}r{{a'}_2})) + 12{{a'}_1}{f_\mathcal{G}}^\prime  - 4{{\rm{e}}^{{a_2}}}{{a'}_1}{f_\mathcal{G}}^\prime \\\nonumber
&- 12{f_\mathcal{T}}{{a'}_2}{f_\mathcal{G}}^\prime  + 4{{\rm{e}}^{{a_2}}}{f_\mathcal{T}}{{a'}_2}{f_\mathcal{G}}^\prime  + 4{{\rm{e}}^{{a_2}}}r{f_\mathcal{R}}^\prime  + 4{{\rm{e}}^{{a_2}}}{f_\mathcal{T}}r{f_\mathcal{R}}^\prime  + {{\rm{e}}^{{a_2}}}{r^2}{{a'}_1}{f_\mathcal{R}}^\prime \\\nonumber
&- {{\rm{e}}^{{a_2}}}{f_\mathcal{T}}{r^2}{{a'}_2}{f_\mathcal{R}}^\prime  + 8{f_\mathcal{T}}{f_\mathcal{G}}^{\prime \prime } - 8{{\rm{e}}^{{a_2}}}{f_\mathcal{T}}{f_\mathcal{G}}^{\prime \prime } + 2{{\rm{e}}^{{a_2}}}{f_\mathcal{T}}{r^2}{f_\mathcal{R}}^{\prime \prime }),\\\nonumber
P_t &= \frac{1}{{4\left( {1 + {f_\mathcal{T}}} \right){r^2}}}{{\rm{e}}^{ - 2{a_2}}}( - {{\rm{e}}^{{a_2}}}{f_\mathcal{R}}( - 4{f_\mathcal{T}} + 4{{\rm{e}}^{{a_2}}}{f_\mathcal{T}} + 2{{\rm{e}}^{{a_2}}}(1\\\nonumber
&+ {f_\mathcal{T}}){r^2}\mathcal{R} - {r^2}{{a'}_1}^2 + 2r{{a'}_2} + 4{f_\mathcal{T}}r{{a'}_2} + r{{a'}_1}\left( { - 2 + r{{a'}_2}} \right)\\\nonumber
&- 2{r^2}{{a''}_1}) + 2({{\rm{e}}^{2{a_2}}}f{r^2} + {{\rm{e}}^{2{a_2}}}f{f_\mathcal{T}}{r^2} - {{\rm{e}}^{2{a_2}}}\left( {1 + {f_\mathcal{T}}} \right){r^2}{f_\mathcal{G}}\mathcal{G} + 2r{{a'}_1}^2{f_\mathcal{G}}^\prime \\\nonumber
&- 12{f_\mathcal{T}}{{a'}_2}{f_\mathcal{G}}^\prime  + 4{{\rm{e}}^{{a_2}}}{f_\mathcal{T}}{{a'}_2}{f_\mathcal{G}}^\prime  + 2{{\rm{e}}^{{a_2}}}r{f_\mathcal{R}}^\prime  + 4{{\rm{e}}^{{a_2}}}{f_\mathcal{T}}r{f_\mathcal{R}}^\prime  - {{\rm{e}}^{{a_2}}}{r^2}{{a'}_2}{f_\mathcal{R}}^\prime \\\nonumber
&- {{\rm{e}}^{{a_2}}}{f_\mathcal{T}}{r^2}{{a'}_2}{f_\mathcal{R}}^\prime  + 4r{f_\mathcal{G}}^\prime {{a''}_1} + 8{f_\mathcal{T}}{f_\mathcal{G}}^{\prime \prime } - 8{{\rm{e}}^{{a_2}}}{f_\mathcal{T}}{f_\mathcal{G}}^{\prime \prime } + r{{a'}_1}( - 6{{a'}_2}{f_\mathcal{G}}^\prime \\
&+ {{\rm{e}}^{{a_2}}}r{f_\mathcal{R}}^\prime  + 4{f_\mathcal{G}}^{\prime \prime }) + 2{{\rm{e}}^{{a_2}}}{r^2}{f_\mathcal{R}}^{\prime \prime } + 2{{\rm{e}}^{{a_2}}}f\mathcal{T}{r^2}{f_\mathcal{R}}^{\prime \prime })).
\end{align}
where
\begin{equation}
\mathcal{R} = \frac{1}{{2{r^2}}}{{\rm{e}}^{ - {a_2}}}\left( {4 - 4{{\rm{e}}^{{a_2}}} + {r^2}{{a'}_1}^2 - 4r{{a'}_2} + r{{a'}_1}\left( {4 - r{{a'}_2}} \right) + 2{r^2}{{a''}_1}} \right)
\end{equation}
\begin{equation}
\mathcal{G} = \frac{{2{{\rm{e}}^{ - 2{a_2}}}}}{{{r^2}}}\left[ {\left( {1 - {{\rm{e}}^{{a_2}}}} \right)2{{a''}_1} + {{a'}_1}{{a'}_2}\left( {{{\rm{e}}^{{a_2}}} - 3} \right) + \left( {1 - {{\rm{e}}^{{a_2}}}} \right){{a'}_1}^2} \right].
\end{equation}
Here prime denotes the derivative with respect to radial coordinate $r$.
\section{Appendix B}
Now by splitting the $f(\mathcal{R,G,T})=f_1(\mathcal{R})+f_2(\mathcal{G})+\lambda \mathcal{T}$, along with the definition of metric variable $a_1$ and $a_2$, we get the solution to field equations as
\begin{align}\nonumber
\rho=&\frac{1}{2 (\lambda +1) (2 \lambda +1) r^2}
(e^{-2 r^2 A_1} (2 e^{r^2 A_1} \lambda  A_2^2
{f_1}' r^4-2 e^{r^2 A_1} \lambda  A_1 A_2\\\nonumber
&{f_1}' r^4+2 e^{r^2 A_1} A_1 \mathcal{R}' {f_1}''
r^3+3 e^{r^2 A_1} \lambda  A_1 \mathcal{R}' {f_1}'' r^3+3
e^{r^2 A_1} \lambda  A_2 \mathcal{R}' {f_1}''\\\nonumber
&r^3+8 \lambda
A_2^2 \mathcal{G}' {f_2}'' r^3-24 \lambda  A_1 A_2 \mathcal{G}'
{f_2}'' r^3-e^{2 r^2 A_1} (\lambda +1)
{f_1} r^2-e^{2 r^2 A_1} (\lambda +1) {f2}\\\nonumber
&r^2+e^{2 r^2 A_1} \mathcal{R} {f1}' r^2+e^{2 r^2 A_1}
\lambda  \mathcal{R} {f_1}' r^2+4 e^{r^2 A_1} A_1
{f_1}' r^2+8 e^{r^2 A_1} \lambda  A_1
{f_1}' r^2+6 e^{r^2 A_1} \lambda  A_2\\\nonumber
&{f_1}' r^2+e^{2 r^2 A_1} \mathcal{G} {f_2}'
r^2+e^{2 r^2 A_1} \lambda  \mathcal{G} {f_2}' r^2+8 \lambda
A_2 {f_2}'' \mathcal{G}'' r^2-2 e^{r^2 A_1}\\\nonumber
&{f_1}'' \mathcal{R}'' r^2-3 e^{r^2 A_1} \lambda
{f_1}'' \mathcal{R}'' r^2-2 e^{r^2 A_1} \mathcal{R}'^2
{f_1}^{(3)} r^2-3 e^{r^2 A_1} \lambda  \mathcal{R}'^2\\\nonumber
&{f_1}^{(3)} r^2+8 \lambda  A_2 \mathcal{G}'^2
{f_2}^{(3)} r^2-4 e^{r^2 A_1} \mathcal{R}'
{f_1}'' r-6 e^{r^2 A_1} \lambda  \mathcal{R}'\\\nonumber
&{f_1}'' r-8 e^{r^2 A_1} A_1 \mathcal{G}' {f_2}''
r-20 e^{r^2 A_1} \lambda  A_1 \mathcal{G}' {f_2}'' r+60
\lambda  A_1 \mathcal{G}' {f2}'' r+24 A_1 \mathcal{G}'\\\nonumber
&{f_2}'' r-4 e^{r^2 A_1} \lambda  A_2 \mathcal{G}'
{f_2}'' r+20 \lambda  A_2 \mathcal{G}'(r) {f2}''
r-2 e^{r^2 A_1} {f_1}'+2 e^{2 r^2 A_1}
{f_1}'-4 e^{r^2 A_1} \lambda  {f_1}'+4\\\nonumber
&e^{2 r^2 A_1} \lambda  {f1}'+8 e^{r^2 A_1}
{f_2}'' \mathcal{G}''+20 e^{r^2 A_1} \lambda
{f_2}'' \mathcal{G}''-20 \lambda  {f_2}''\\\nonumber
&\mathcal{G}''-8 {f_2}'' \mathcal{G}''+8 e^{r^2 A_1} \mathcal{G}'^2
{f_2}^{(3)}+20 e^{r^2 A_1} \lambda  \mathcal{G}'^2
{f_2}^{(3)}-20 \lambda  \mathcal{G}'^2\\\nonumber
&{f_2}^{(3)}-8 \mathcal{G}'^2 {f_2}^{(3)}))
\end{align}

\begin{align}\nonumber
P_r&=\frac{1}{2 (\lambda +1) (2 \lambda +1) r^2}
(e^{-2 A_1 r^2}(-\lambda  r^2 e^{A_1 r^2}
   {f_1}^{(3)} \mathcal{R}'^2-\lambda  r^2 e^{A_1 r^2} \mathcal{R}''\\\nonumber
&   {f_1}''+6 \lambda  r e^{A_1 r^2} \mathcal{R}'
   {f_1}''+4 r e^{A_1 r^2} \mathcal{R}' {f_1}''+A_1
   \lambda  r^3 e^{A_1 r^2} \mathcal{R}' {f_1}''+A_2 \lambda\\\nonumber
&   r^3 e^{A_1 r^2} \mathcal{R}' {f_1}''+2 A_2 r^3 e^{A_1 r^2}
   \mathcal{R}'(r) {f_1}''-\lambda  r^2 e^{2 A_1 r^2} \mathcal{R}
   {f_1}'+2 A_2 \lambda  r^2 e^{A_1 r^2}\\\nonumber
&   {f_1}'+4 \lambda  e^{A_1 r^2} {f_1}'-4
   \lambda  e^{2 A_1 r^2} {f_1}'-r^2 e^{2 A_1 r^2} \mathcal{R}
   {f_1}'+4 A_2 r^2 e^{A_1 r^2} {f_1}'+2
   e^{A_1 r^2} {f_1}'-2 e^{2 A_1 r^2}\\\nonumber
&   {f_1}'-2 A_2^2 \lambda  r^4 e^{A_1 r^2}
   {f_1}'+2 A_1 A_2 \lambda  r^4 e^{A_1 r^2}
   {f_1}'+(\lambda +1) r^2 e^{2 A_1 r^2}\\\nonumber
&   {f_1}-8 A_2 \lambda  r^2 {f_2}^{(3)}
   \mathcal{G}'^2-4 \lambda  e^{A_1 r^2} {f_2}^{(3)} \mathcal{G}'^2-8
   A_2 \lambda  r^2 \mathcal{G}'' {f_2}''-4 \lambda  \\\nonumber
& e^{A_1 r^2} \mathcal{G}'' {f_2}''-8 A_2^2 \lambda  r^3 \mathcal{G}'
   {f_2}''+24 A_1 A_2 \lambda  r^3 \mathcal{G}'
   {f_2}''+4 A_1 \lambda  r e^{A_1 r^2} \mathcal{G}'\\\nonumber
&   {f_2}''-12 A_2 \lambda  r e^{A_1 r^2} \mathcal{G}'
   {f_2}''-8 A_2 r e^{A_1 r^2} \mathcal{G}'
   {f_2}''-12 A_1 \lambda  r \mathcal{G}' {f_2}''+28\\\nonumber
&   A_2 \lambda  r \mathcal{G}' {f_2}''+24 A_2 r \mathcal{G}'
   {f_2}''-\lambda  r^2 e^{2 A_1 r^2} \mathcal{G}
   {f_2}'-r^2 e^{2 A_1 r^2} \mathcal{G}\\\nonumber
&   {f_2}'+(\lambda +1) r^2 e^{2 A_1 r^2}
   {f_2}+4 \lambda  {f_2}^{(3)} \mathcal{G}'^2+4
   \lambda  \mathcal{G}'' {f_2}''))
\end{align}

\begin{align}\nonumber
P_t&=\frac{1}{2 (\lambda +1) (2 \lambda +1) r^2}
(2 e^{r^2 A_1} A_2^2 {f_1}' r^4+2 e^{r^2 A_1} \lambda
   A_2^2 {f_1}' r^4-2 e^{r^2 A_1} A_1 A_2\\\nonumber
&   {f_1}' r^4-2 e^{r^2 A_1} \lambda  A_1 A_2
   {f_1}' r^4-2 e^{r^2 A_1} A_1 \mathcal{R}' {f_1}''
   r^3-3 e^{r^2 A_1} \lambda  A_1 \mathcal{R}' {f_1}'' r^3+2\\\nonumber
&   e^{r^2 A_1} A_2 \mathcal{R}'(r) {f_1}'' r^3+e^{r^2 A_1}
   \lambda  A_2 \mathcal{R}' {f_1}'' r^3+8 \lambda  A_2^2
   \mathcal{G}' {f_2}'' r^3+8 A_2^2 \mathcal{G}' {f_2}''\\\nonumber
&   r^3-24 \lambda  A_1 A_2 \mathcal{G}' {f_2}'' r^3-24 A_1 A_2
   \mathcal{G}' {f_2}'' r^3+e^{2 r^2 A_1} (\lambda +1)
   {f_1} r^2+e^{2 r^2 A_1} (\lambda +1) {f_2}\\\nonumber
&   r^2-e^{2 r^2 A_1} \mathcal{R} {f_1}' r^2-e^{2 r^2 A_1}
   \lambda  \mathcal{R} {f_1}' r^2-2 e^{r^2 A_1} A_1
   {f_1}' r^2-4 e^{r^2 A_1} \lambda  A_1\\\nonumber
&   {f_1}' r^2+4 e^{r^2 A_1} A_2 {f_1}' r^2+2
   e^{r^2 A_1} \lambda  A_2 {f_1}' r^2-e^{2 r^2 A_1}
   \mathcal{G} {f_2}' r^2-e^{2 r^2 A_1} \lambda  \mathcal{G}\\\nonumber
&   {f_2}' r^2+8 \lambda  A_2 {f_2}'' \mathcal{G}''
   r^2+8 A_2 {f_2}'' \mathcal{G}'' r^2+2 e^{r^2 A_1}
   {f_1}'' \mathcal{R}'' r^2+3 e^{r^2 A_1} \lambda\\\nonumber
&   {f_1}'' \mathcal{R}'' r^2+2 e^{r^2 A_1} \mathcal{R}'^2
   {f_1}^{(3)} r^2+3 e^{r^2 A_1} \lambda  \mathcal{R}'^2
   {f_1}^{(3)} r^2+8 \lambda  A_2 \mathcal{G}'^2\\\nonumber
&   {f_2}^{(3)} r^2+8 A_2 \mathcal{G}'^2 {f_2}^{(3)}
   r^2+2 e^{r^2 A_1} \mathcal{R}' {f_1}'' r+2 e^{r^2 A_1}
   \lambda  \mathcal{R}' {f_1}'' r+4 e^{r^2 A_1} \lambda  A_1\\\nonumber
&   \mathcal{G}' {f_2}'' r-12 \lambda  A_1 \mathcal{G}'
   {f_2}'' r+4 e^{r^2 A_1} \lambda  A_2 \mathcal{G}'
   {f_2}'' r-4 \lambda  A_2 \mathcal{G}' {f_2}'' r+8\\\nonumber
&   A_2 \mathcal{G}' {f_2}'' r-4 e^{r^2 A_1} \lambda
   {f_2}'' \mathcal{G}''(r)+4 \lambda  {f_2}'' \mathcal{G}''-4
   e^{r^2 A_1} \lambda  \mathcal{G}'^2 {f_2}^{(3)}+4 \lambda
   \mathcal{G}'^2 {f_2}^{(3)})
\end{align}
Here, $A_1$, $A_2$ and $A_3$ are constant parameters for compact stars.

\end{document}